\documentclass[12pt]{article}

\usepackage{graphicx}
\usepackage{amssymb}
\usepackage{amsthm}
\usepackage{booktabs}
\usepackage{rotating}
\usepackage{multirow}
\usepackage{eurosym}
\usepackage{amsfonts}
\usepackage{amsmath}
\usepackage{umoline}
\usepackage{caption}
\usepackage{color}
\usepackage{url}
\usepackage{authblk}
\usepackage[english]{babel}

\usepackage{pbox}
\usepackage{footnote}
\usepackage{multirow}
\usepackage{changepage}
\usepackage{tablefootnote}
\usepackage{ccicons}

\usepackage[margin = 2cm]{geometry}

\usepackage{hyperref}

\usepackage[natbibapa]{apacite}
\usepackage[symbol]{footmisc}
 
\usepackage{tikz} 
\usepackage{mwe} 
\usepackage{subfigure}
\usepackage{xcolor}
\usepackage[utf8]{inputenc}
\usepackage{import}
\usepackage{float} 
\usepackage{comment}
\usepackage{array}
\newcolumntype{P}[1]{>{\centering\arraybackslash}p{#1}}
\newcolumntype{M}[1]{>{\centering\arraybackslash}m{#1}}
\usepackage{tabularx}

\begin{document}

\title{On the combination of graph data for assessing thin-file borrowers' creditworthiness}\footnotetext[1]{\scriptsize NOTICE: This is a preprint of a published work. Changes resulting from the publishing process, such as editing, corrections, structural formatting, and other quality control mechanisms may not be reflected in this version of the document. Please cite this work as follows: Muñoz-Cancino, R., Bravo, C., R\'{i}os, S. A., \& Graña, M. (2022). On the combination of graph data for assessing thin-file borrowers’ creditworthiness. Expert Systems with Applications, 118809. DOI: https://doi.org/10.1016/j.eswa.2022.11880}

\author[1]{Ricardo Muñoz-Cancino}
\author[2]{Cristi\'{a}n Bravo}
\author[3]{Sebasti\'{a}n A. R\'{i}os}
\author[4]{Manuel Graña}

\affil[1,3]{Business Intelligence Research Center (CEINE), Industrial Engineering Department, University of Chile, Beauchef 851, Santiago 8370456, Chile}
\affil[2]{Department of Statistical and Actuarial Sciences, The University of Western Ontario, 1151 Richmond Street, London, Ontario, N6A 3K7, Canada.}
\affil[4]{Computational Intelligence Group, University of Basque Country, 20018 San Sebasti\'{a}n, Spain.}

\date{}

\maketitle

\begin{abstract}
    Thin-file borrowers are customers for whom a creditworthiness assessment is uncertain due to their lack of credit history. To address missing credit information,  many researchers have used borrowers' social interactions as an alternative data source. Exploiting social networking data has traditionally been achieved by hand-crafted feature engineering, but lately, graph neural networks have emerged as a promising alternative.
    Here we introduce an information-processing framework to improve credit scoring models by blending several methods of graph representation learning: feature engineering, graph embeddings, and graph neural networks. In this approach, we aggregate the methods' outputs to be fed to a gradient boosting classifier to produce a final creditworthiness score. We have validated this framework over a unique multi-source dataset that characterizes the relationships, interactions, and credit history for the entire population of a Latin American country, applying it to credit risk models, application, and behavior. It also allows us to study both individuals and companies.
    Our results show that the methods of  graph representation learning should be used as complements; they should not be seen as self-sufficient methods, as it is currently done. We improve the creditworthiness assessment performance in terms of the measures of Area Under the ROC Curve (AUC) and Kolmogorov-Smirnov (KS), outperforming traditional methods of exploiting social interaction data.
    In the area of corporate lending, where the potential gain is much higher, our results confirm that the evaluation of a thin-file company cannot solely consider the company’s own  characteristics. The business ecosystem in which these companies interact with their owners, suppliers, customers, and other companies provides novel knowledge that enables financial institutions to enhance their creditworthiness assessment. 
    Our results let us know when and on which population to use graph data and the expected effects on performance. They also show the enormous value of graph data on the credit scoring problem for thin-file borrowers, mainly to help companies with thin or no credit history to enter the financial system.

\end{abstract}


\section{Introduction}

A large part of the population requires access to credit to achieve their life goals: social mobility, owning a home, and financial success. Moreover, access to financial services and a proper credit evaluation can facilitate and are often necessary to obtain a job, rent a home, buy a car, start a new business, or pursue a college education \citep{Hurley2017, Aziz2019}. 
At the macroeconomic level, access to credit is a major driver for local economic growth, especially in developing economies \citep{Diallo2017}.
Financial institutions play a significant social role in facilitating access to credit and facing the  entailed risks of lending money. To manage this credit risk, financial institutions have applied credit scoring models to assess the creditworthiness of their borrowers, that is, to distinguish between good and bad payers and delivering loans to those who are most likely to repay. To build a credit scoring model, financial institutions often use personal information, banking data, and payment history to estimate creditworthiness and the probability of default.
Despite being the standard mechanism in the industry for credit-granting decisions and  the management of the loan's life cycle \citep{Thomas2017}, this ubiquitous tool still does not ensure adequate access to credit and to the financial system.

The World Bank estimates that more than 1.4 billion adults remain unbanked,  without access to the financial system \citep{Findex2022}. This number only considers those who do not have a bank account through either a financial institution or mobile banking. If we included  underbanked people, that is, those who have an account but cannot apply for a loan, this number would be much larger. 
Being unbanked or underbanked raises the issue of those who lack a credit history, also known as thin-file borrowers: people who have no access to a loan not because they are bad payers but because they lack the attributes evaluated by traditional credit scoring models \citep{OECD2013, Hurley2017, Baidoo2020, Djeundjie2021}.

In this scenario, lenders have tried different ways to reach this population; we  highlight two business models here. In the traditional business model, the higher risk assumed due to the lack of information is compensated by applying higher interest rates. Alternatively,  granting microcredits has been used as a strategy to assess the client's payment behavior under limited exposure. However, neither of these solutions has proven to be cost-effective in addressing the credit needs of this population \citep{Hurley2017, Baidoo2020}.

For this reason, financial institutions, fintech, and researchers have looked in recent years for business-model innovations and better decision-making with the available information. This search is done via developing better scoring algorithms and using alternative data sources to improve credit scoring models. Regarding the use of alternative information, graph data has gained high visibility because it allows an improvement of credit scoring models' performance \citep{Oskarsdottir2019, Roa2021}.

We have  identified two main gaps, which are addressed in this work. The first gap is the data sources employed. Most of the studies are carried out with partial social networks that fail to capture the overall picture of the client's social interactions. These networks are limited by the data provider. Our study  uses social networking data  to characterize the interactions of the country's entire population, encompassing the complete financial system.
Secondly, the network knowledge extraction is mainly done both through hand-made feature engineering \citep{Freedman2017, Ruiz2018, Oskarsdottir2019, Niu2019} and, in recent years, through graph neural networks \citep{Roa2021} that do are no improvement over the traditional feature-engineering approach.

Our work will investigate the combination of different representation learning techniques with complex graph structures instead of observing them in isolation. Hence, we formulate the following research questions:

\begin{enumerate}
    \item When combining different graph representation learning (GRL) techniques over complex graph structures, is there a performance improvement compared to merely applying hand-crafted feature engineering or graph neural networks?
    \item What insights are obtained into the combined network features, and what value do these insights add to credit risk assessment?
    \item Where does social information help the most? Is the most significant performance enhancement obtained in  personal credit scoring or business credit scoring? What can we gather from this information? Does it influence which network and which features are the most relevant?
\end{enumerate}

This study challenges traditional hand-crafted feature engineering and the novel approach of graph neural networks (GNNs) by combining multiple GRL methods. In particular, our work contributes to the following aspects.

\begin{itemize}
    \item We introduce a framework to combine traditional hand-engineered features with graph embeddings and GNN features. This framework produces a single score, helping its users decide whether to approve or reject a credit.
    \item Our results are the first to validate and test graph data regarding both corporate and consumer lending, showing that the information from graphs has a different effect depending on the analyzed borrower, people, or companies. These effects are reflected both in the predictive power enhancement and in the features relevant in each problem, letting us know not only when and on which population to use social-interaction data but also which effects on creditworthiness prediction performance to expect.
    \item To the best of our knowledge, this is the first study that considers the credit behavior of an entire country, together with social networks that allow the characterizing of its entire population and consolidate multiple types of social and economic relationships, for example, parental, spouses, business owners, employers and employees, or transactional services.
    \item This paper also contributes to the growing literature in credit scoring and network data, proposing a mechanism to achieve better results than the popular hand-crafted feature engineering and the novel GNN approach.
\end{itemize}

This paper is structured as follows. Section \ref{sec:relatedWork} presents a review of credit risk management, credit scoring and social networks. The GRL methods are presented in Section \ref{sec:ReprLearning}.
Section \ref{sec:datadescr} describes the data sources and features extracted for classification.
Section \ref{sec:ExpDesign} shows the proposed information-processing methodology and the adopted experimental design. Section \ref{sec:Results} presents the results obtained. The conclusions and future work that originated from this research are presented in Section \ref{sec:Conclusions}.

\section{Background and Related Work}\label{sec:relatedWork}
\subsection{Credit Risk Management}

Banks' core business is  granting loans to individuals and companies. Granting a loan is not risk-free; in fact, banks are heavily exposed to credit risk \citep{Anderson2022}, originating from the potential loss due to the debtors' default or their inability to comply with the agreed conditions \citep{Basel2000}. Banking risk management focuses on detecting, measuring, reporting, and managing all sources of risk. Banks define strategies, policies, and procedures to limit the assumed risk. These strategies encourage and integrate the use of mathematical models for the early detection of potential risks.
Credit scoring is  widely used  for managing credit risk, handling large volumes of data, and capturing complex patterns that are difficult  to express as simple business rules. This instrument became popular and ubiquitous in the 1980s, mainly due to advances in computing power and to the growth of financial markets, which made it almost impossible to manage large credit portfolios without this kind of tool \citep{Thomas2017}.

The regulatory framework also endorses the use of credit scoring models; in fact, the Basel Accords allow banks to manage credit risk with internal ratings. Specifically, banks develop internal models for assessing the expected loss. This assessment can be divided into three components: the probability of default (PD), the loss given default (LGD), and the exposure at default (EAD). The PD is a key component, because it is used to define the credit granting  policies and for portfolio management. The general approach to estimating the PD and assessing the borrower's creditworthiness is through classification techniques using demographic features and payment history as explanatory variables. 

Over the years, lenders have explored multiple ways to improve creditworthiness assessment, novel machine learning techniques \citep{Moscato2021}, and non-traditional data sources \citep{Aziz2019}. Multiple lines of research have been established; some of them attempt to understand the characteristics of defaulters \citep{Bravo2015}, the feature selection process \citep{Kozodoi2019, Maldonado2017}, or the transformation of the feature space \citep{Carta2021}. However, the most significant improvements have been obtained by the exploitation of alternative data sources such as telephone call data \citep{Oskarsdottir2017, Oskarsdottir2018, Oskarsdottir2019}, written risk assessments \citep{Stevenson2020}, data generated by an app-based marketplace \citep{Roa2021eswa, Roa2021}, social media data \citep{Tan2018, Cnudde2019, Septian2020}, network information \citep{Ruiz2018}, behavioral and psychological surveys \citep{Goel2021}, fund transfers datasets \citep{Shumovskaia2020, Sukharev2020}, and psychometric data \citep{Rabecca2018, Djeundjie2021, Rathi2022}. All these studies have in common the use of social-interaction information, the graph formed of the interactions among individuals recorded in alternative data sources.

There are multiple taxonomies of credit scoring problems. One that has been widely adopted by academics and practitioners   distinguishes between application scoring and behavior scoring. On the one hand,   application scoring corresponds to a credit scoring system for new customers, where the available information is often scarce and limited. On the other hand, behavioral scoring is  a credit scoring system for borrowers with available credit and repayment history. In the current study, both of these credit scoring types and their differences between personal and business clients are explored.

\subsection{Credit Risk and Social Networks}

The inclusion of alternative data in the credit scoring problem has gained relevance in recent years. We define \textit{graph data} as any information that records the relationships or interactions among entities that can be represented by a set of nodes in which edges connect pairs of nodes. We refer to a network as a \textit{Social Network} when nodes are people or companies, and edges denote any social interaction, such as among friends, acquaintances, neighbors, colleagues, or affiliations with the same group \citep{Romero2019}. Mathematically, we describe a network through a graph $G(V,E,A)$, where $V$ is the set of nodes, and $E$ is the set of edges. Let $V=\{v_1, \ldots, v_N\}$ where $|V| = N$ is the number of nodes, and the adjacency matrix  $A \in \mathbb{R}^{|V| \times |V|}$ with $A_{ij} = 1$ if there is an edge $e_{ij}$ from $v_i$ to $v_j$, $A_{ij} = 0$ otherwise. Additionally, the graph can be associated with a  matrix of node attributes $X \in \mathbb{R}^{N \times F}$, where $X_i \in \mathbb{R}^{F} $ represents the feature vector of node $v_i$.

The emerging literature on credit scoring and network data has focused on incorporating hand-crafted features into a traditional credit scoring problem \citep{Oskarsdottir2017, Oskarsdottir2017b}. The authors incorporate network information into the formulation of the customer churn problem, using eight telco datasets originating from around the world. This series of studies outlines the foundations for the incorporation of network data in credit scoring. This framework is applied to the credit scoring problem by \citet{Oskarsdottir2018, Oskardottir2018b, Oskarsdottir2019}, where the authors introduce a methodology to enhance smartphone-based credit scoring models' predictive power through feature engineering from a pseudo-social network, combining social-network analysis and representation learning. According to this research, it is feasible to increase the performance of micro-lending smartphone applications, generating high helping potential for financial inclusion.

An extension of this work is to measure the temporal and topological dynamics of credit risk, that is, how it evolves and spreads over the graph representation of the social network. For instance, \citet{Bravo2020} implemented modifications to the PageRank algorithm to quantify this phenomenon. 
This methodology allows them to quantify the risk of the different entities in a multilayer network. Their results show how the risk of default spreads and evolves over a network of agricultural loans. Then, \citet{Oskarsdottir2020} analyzed how to build the multilayer network, interpret the variables derived from it, and incorporate this knowledge into credit risk management. Their results reveal that the default risk increases as a debtor presents links with many defaulters; however, this effect is mitigated by the size of each individual's neighborhood. These results are significant because they indicate that default and financial-stability risk spread through the network.

Other works have used an approach based on a graph convolutional network (GCN) for this purpose. \citet{Shumovskaia2020} present one of the first empirical works with massive graphs created from transactions between clients of a large Russian bank. They propose a framework to estimate links using SEAL \citep{ZhangCheng2018} and recurrent neural networks, the SEAL-RNN framework. One of the advantages of using SEAL is that it focuses only on the link's neighborhood to be predicted, and it does not use the entire graph as in GCN. This framework permits the analysis to be scaled to massive graphs of 86 million nodes and 4 billion edges. Although the framework is not a methodology for default prediction, \citeauthor{Shumovskaia2020} extend the scope of their research and apply it to a credit scoring problem. \citet{Sukharev2020} propose a method to predict the default from a money transfer network and the historical information of transactions. To work with both datasets, they propose a methodology based on GCN  and recurrent neural networks to handle network data and transactional data, respectively. As baseline models, they train a model with 7000 features; however, they achieve an increase of 0.4\% AUC when comparing the proposed model with the best baseline model. Finally, \citet{Roa2021eswa} present a methodology for using alternative information in a credit scoring model. Models are estimated using data generated by an app-based marketplace. This information is precious for low-income segments and young individuals, who are often not assessed well by traditional credit scoring models. The authors compare a model with hand-crafted features versus models from GCNs. However,   GCNs do not achieve better results than do hand-crafted features in terms of predictive power.

\section{Representation Learning on Networks}\label{sec:ReprLearning}
The machine learning subfield that works on graph-structured data is known as graph representation learning or GRL \citep{Hamilton2017}. Unlike the traditional tabular data,  network data imposes a challenge to conventional machine learning algorithms because it is not possible to use them directly, forcing  changes either on the algorithms or on the data representation. These challenges are required because the network information is, in essence, unstructured. In fact, operations that are easy to calculate on other data types, such as convolutions on images, cannot be applied directly to graphs because each node has a variable number of neighbors.
Researchers have proposed many methodologies to extract knowledge for networks; here, we present a nomenclature and the characteristics of the most popular  methods.

\subsection{Feature Engineering}\label{sec:featEngDesc}

Data preparation is one of the most critical steps in any analytical project before training any machine learning model. Formulating accurate and relevant features is critical to improving model performance \citep{Nargesian2017}.
Regarding the use of graph data, the traditional feature engineering approach consists of characterizing each node either based on the aggregation of its neighborhood's features or the node's statistics within the network. 

\subsection{Network Embeddings}\label{sec:NetEmbeddingTechniques}
Network embedding methods are  unsupervised learning techniques aiming to learn a Euclidean representation of networks in a much lower dimension. Each node is mapped into a Euclidean space through the optimization of similarity functions. The distance between  network nodes in the new space is a surrogate for   the node's closeness within the network  structure. Node embedding techniques often  replace  feature engineering processes.
 
Formally, a  network is represented by a graph  $G(V,E,A)$ defined by a set of nodes $V$, a set of edges $E$, and an adjacency matrix $A \in \mathbb{R}^{|V| \times |V|}$. The embedding of a node is a function $ f: G(V,E,A) \rightarrow \mathbb{R}^d$ that maps each node $v \in V$ to an embedding vector $\{Z_v\} _{v \in V} \in \mathbb{R}^d$ \citep{Arsov2019}, preserving the adjacency in the graph. The embedding vectors of pairs of nodes that are connected by an edge are closer than  those that are disconnected. Let $ Z \in \mathbb{R}^{|V| \times d}$ denote the node-embedding matrix, where $d \ll |V|$ for scalability purposes. 
The most popular network embedding method is Node2vec \citep{Grover2016}, which is an algorithmic framework for learning low-dimensional network representation.  This algorithm maximizes the probability of preserving the neighborhood of the nodes in the embedding subspace.
The algorithm optimizes using stochastic gradient descent, a network-based objective function, and  produces samples for neighborhoods of nodes through second-order random walks. The key feature of Node2vec is the use of biased-random walks,  providing a trade-off among two network search methods: breadth-first search (BFS) and depth-first search (DFS). This trade-off creates more informative network embeddings than other competing methods.

\subsection{Graph  Neural Networks (GNN)}\label{sec:GCN}

Graph-structured data has arbitrary structures that can vary significantly between networks or within different nodes of the same network. Their support domain  is not a uniformly discretized Euclidean space. For this reason, the convolution operator that is often used for signal processing  cannot be directly applied  to graph-structured data.  Geometric deep learning (GDL) and graph neural networks (GNN) aim to modify, adapt and create deep learning techniques for non-Euclidean data.  The proposed GDL computational schemes are an adaptation of deep autoencoders, convolutional networks, and recurrent networks to this particular data domain. In this study, we will be applying graph convolutional networks (GCN) and graph autoencoders (GAE).

\subsubsection{Graph Convolutional Networks}
The Graph convolutional networks generalize the convolution operation to  networks formalized as graphs. The GCNs aim to produce a node's representation $Z_v$ by adding its attributes or feature vector $X_v$ and neighbors $\{X_u\}_{u \in N(v)}$, where $N(v)$ is the ego network of node $n$, that is, the subgraph composed of the nodes to whom node $n$ is connected. This study uses the spectral-based GCN, also known as the Chebyshev spectral convolutional neural network, proposed by \citet{Defferrard2016}, which defines the graph convolution operator   as a filter from graph signal processing. In particular, we use the specific  GCN proposed by \citet{Kipf2016}, which uses as a filter a first-order approximation of the Chebyshev polynomial of the eigenvalues' diagonal matrix. This graph convolution operator follows the expression: 

\begin{equation}\label{eq:GCN1}
    X_i \ast g_\theta  = \theta_0 X_i - \theta_1 D^{-\frac{1}{2}} A D^{-\frac{1}{2}} X_i,
\end{equation}
where  $X_i$ is the feature vector, $g_\theta$ is a function of the eigenvalues of the normalized graph, Laplacian matrix $L = \mathcal{I}_n - D^{-\frac{1}{2}} A D^{-\frac{1}{2}}$, $A$ is the adjacency matrix, $D_{ij} = \sum_j A_{i,j}, \; \forall i,j \in V: i = j$, and $\theta$ is the vector of the Chebyshev coefficients. 
In the following, section we summarize the derivation of this first-order approximation \citep{Kipf2016}.

\subsubsection{Derivation of GCN from Spectral Methods}
 Spectral methods are founded on a solid theoretical basis defined for methods of graph signal processing developed essentially from the Laplacian matrix properties. To build up the graph convolution operator, we start from the normalized graph Laplacian  matrix defined as follows: 
  \begin{equation}
  L = \mathcal{I}_n - D^{-\frac{1}{2}} A D^{-\frac{1}{2}}, 
  \end{equation}
  where $A$ is the adjacency matrix, and $D_{ij} = \sum_j A_{i,j}, \; \forall i,j \in V: i = j$. 
 Because $L$ is  a real, symmetric, and positive semi-defined matrix, 
   we can rewrite L as a function of its eigenvector matrix $U$ and its eigenvalues $\lambda_i$, that is, $L = U \Lambda U^T$,  where $U \in \mathbb{R}^{N \times N}$, and $\Lambda_{ij} = \lambda_i,  \; \forall i,j \in V: i = j$. 
   
   The next step  is   to define the graph Fourier transform and its inverse. The graph Fourier transform $\mathcal{F}$ of the feature vector $X_i \in X$ is defined as follows:
   \begin{equation}
       \mathcal{F}(X_i) = U^T X_i,
   \end{equation}
   where $X$ is the matrix of  node attributes, and the inverse Fourier transform of a graph is defined as follows:
   \begin{equation}
       \mathcal{F}^{-1}(\hat{X_i}) = U \hat{X_i},
   \end{equation}
 where $\hat{X_i}$ are the coordinates of the nodes in the new space. Therefore, the feature vector can be written as $X_i = \sum_{j in V} \hat{X_i} u_j$.
 Finally, the graph convolution of feature vector $X_i$ with filter $g \in  \mathbb{R}^N$, using the element-wise product $\odot$,  is defined as follows:

 \begin{equation}\label{eq:ChebNet}
     X_i \ast g = \mathcal{F}^{-1}(\mathcal{F}(X_i) \odot \mathcal{F}(g))
 \end{equation}

 One of the most popular filters is the Chebyshev polynomial of the eigenvalues' diagonal matrix, that is, $g_\theta = diag(U^Tg) = \sum_K \theta_k T_k (\hat{\Lambda})$, where $\hat{\Lambda} = 2\lambda / \lambda_{max} - \mathcal{I}$ and the polynomials $T_k$ are defined as $T_k(x) = 2xT_{k-1} - T_{k-2}(x)$, with $T_0(x) = 1$ and $T_1(x) = x$. Therefore, the GCN, defined as  the Chebyshev Spectral CNN \citep{Defferrard2016}, takes the following form:

 \begin{equation}
     X_i \ast g_\theta = \sum_K \theta_k T_k (\hat{L})X_i,
 \end{equation}
where $\hat{L} = 2L/\lambda_{max} - \mathcal{I}$. Despite being a graph convolution simplification, this convolution is computationally expensive for large graphs. To solve this problem, \citet{Kipf2016} present a first-order approximation of the Chebyshev Spectral CNN. Assuming $K=1$ and $\lambda_{max}=2$ , the Equation \ref{eq:ChebNet} takes the following form: 

 \begin{equation}\label{eq:GCN}
     X_i \ast g_\theta  = \theta_0 X_i - \theta_1 D^{-\frac{1}{2}} A D^{-\frac{1}{2}} X_i
 \end{equation}

\subsubsection{Graph Autoencoders (GAEs)}\label{sec:GAE}

Graph autoencoders (GAEs) are an unsupervised method to obtain a low-dimensional representation of the network. The objective of the GAE is to reconstruct the original network using the same network for this task but encoding it, reducing its dimensionality, and then decoding it to reconstruct the network.
The encoded representation is used as the network embedding. \citet{Wu2019} distinguish two main uses of GAEs, namely graph generation and network embedding. This research will use GAEs to obtain a lower-dimensional vector, preserving the network topology (network embedding). 

The previously defined GCN  is the building block of the GAE architecture and allows the simultaneous encoding of the network topology and the attributes of the nodes. The GAE \citep{Kipf2016VGAE} calculates the network embedding matrix $Z$ and the reconstruction of the original network adjacency matrix  $\hat{A}$ as follows:

\begin{equation}
    \hat{A} = \sigma(ZZ^T), \; with \; Z = GCN(X,A),
\end{equation}
where $X$ is the matrix of node attributes, and $A$ is the network's adjacency matrix.

\section{Data Description}\label{sec:datadescr}

The data used in this paper encompasses several datasets provided by  a large Latin American bank. Some datasets contain information from their customers, while others concern the entire population of the country. 

\subsection{Ethical and Privacy Protection Considerations}
The datasets contain anonymized information and do not compromise the identity of any customer or their personal information in any way. The datasets share an anonymized customer's ID allowing us to merge multiple sources. 
Regarding the value, importance, and sensitivity of the data, we have applied multiple actions to ensure its security, integrity, and confidentiality. Customer identifiers and any personal data were removed before starting the analysis, and there is no possibility that this investigation can leak any personal private information. In addition, any final data produced as a result of this research does not compromise customers' privacy.

\subsection{Social-Interaction Data}\label{sec:networkdata}

The information collected by the financial institution to construct the social network background information of the thin file borrower originates from varied sources and can be cataloged as follows:

\begin{itemize}
    \item \textbf{[WeddNet] Network of marriages:} This network is built from the information of marriages recorded by the bureau of vital statistics from 1938 to December 2015.  It includes the anonymized identifiers of the husband and the wife and the wedding date.
    \item \textbf{[TrxSNet] Transactional services network:} The primary source of this network comes from transactional services data, primarily payroll services and the transfers of funds between two entities. We have access to monthly data from January 2017 to December 2019.
    \item \textbf{[EnOwNet] Enterprise's ownership network:} This network is built from the information on companies' ownership structure. For each firm, we have information concerning their owners, be they individuals or other firms. We have quarterly information from January 2017 to November 2019.
    \item \textbf{[PChNet] Parents and children network:} This network corresponds to parental relationships. For every person born between January 1930 and June 2018, we have the anonymized identifiers of their parents.
    \item \textbf{[EmpNet] Employment network:} This network is built from multiple sources and connects people with their employers. We have monthly data from January 2017 to December 2019.
\end{itemize}

\subsection{Financial Data}\label{sec:nodedata}
The node dataset contains information on the consolidated indebtedness of each debtor in the financial system from January 2018 until March 2020, reporting monthly the debt decomposition from 7.65 million people and 245,000 firms. We refer to the features extracted from this dataset as node features. Additionally, for every person and firm in the previous dataset, we have access to the BenchScore, which corresponds to the probability of default for the coming 12 months. This probability was assessed and provided by the financial institution, and it is our benchmark to contrast the performance of our models.

\subsection{Network Construction}

It is possible to build a network from each of the data sources indicated in Section \ref{sec:networkdata}. However, they share characteristics that allow them to be grouped. For this reason, we combine the networks into two primary data sources, from which we construct networks that characterize people and businesses.

\textbf{[FamilyNet] Family network:} This network is formed through the combination of the network of marriages (WeddNet) and the parent and children network (PChNet). For the construction of this network, we use the historical information available until the beginning of the analysis period; no further information is included. In this way, the network remains unvarying throughout the study. We call this type of network a static network.

 \textbf{[EOWNet] Enterprise's ownership Network and Workers:} This network is composed of the fusion of the transactional services network (TrxSNet), the enterprise's ownership network (EnOwNet), and the employment network (EmpNet). This network attempts to represent the business ecosystem in which companies, business owners, and employees interact.
Based on these data sources, a series of 24 networks are generated, one for each of the 24 months available, including the information collected up to the last day of the corresponding month. We call this type of network a  temporal network, because nodes and edges change over time. 
 
\section{Experimental Design and Methodology}\label{sec:ExpDesign}

\subsection{Datasets}\label{sec:Rdataset}
The credit scoring models are built with information about the financial system for 24 months. However, the models are trained over 23 months because the first month  is left out for the feature extraction process (presented in  Section \ref{sec:featEng})  in order to avoid target leakage \citep{Kaufman2011}.
For the unbanked application scoring model, individuals and companies are considered only in the month that they enter the financial system. In contrast, for the behavioral scoring model individuals and companies are considered six or more months after entering the financial system. Table \ref{table:datasetdesc}  summarizes the scoring application, the model trained and the size of the dataset used.

\begin{table}[ht]
\footnotesize
\centering
\caption{Description of dataset}
\label{table:datasetdesc}
\noindent 
\begin{tabular}{|p{4cm}|p{4cm}|c c|}
\hline
{\centering  Scoring application} &  {\centering Model} & {\centering Observations} & {\centering \# Features} \\ \hline
\multirow{2}{*}{\parbox{4cm}{\centering Unbanked Application Scoring}} &  Business Credit Score & 29,044 & 687 \\ 
&   Personal Credit Score & 192,942 & 1,283\\ \cline{1-4}
\multirow{2}{*}{\parbox{4cm}{\centering Behavioral Scoring}} &  Business Credit Score & 931,910 & 687 \\               
& Personal Credit Score &  1,978,664 & 1,283\\ \hline
\end{tabular}
\end{table}

\subsection{Target}\label{sec:Rtarget}
The target event was "becoming a defaulter during the period of observation". Therefore, we only took into account individuals or businesses that were non-defaulters at the start of the period of observation; we dismissed entities that were defaulters at the very beginning of the observation.
In the current study, a person or company was considered a defaulter when they had payments past the due date for 90 or more days  within  12 months starting from the observation point. Otherwise, they were considered non-defaulters. The target vector, denoted by $y_{def}$, contained the actual information about the target event.

\subsection{Traditional and Graph Representation Learning Features}\label{sec:featEng}

Combining the node information from Section \ref{sec:nodedata} and the network data from Section \ref{sec:networkdata} makes it possible to generate a set of new characteristics through a feature extraction process. The sets of characteristics generated are detailed below:

\begin{itemize}
    \item \textbf{[NodeStats] Node Statistics:} This dataset collects node centrality statistics, namely, its degree, degree centrality, number of triads, PageRank score, authority and hub score given by the Hits algorithm \citep{Kleinberg-1999}, and an indicator of whether the node is an articulation point.
    
    \item \textbf{[EgoNet] EgoNetwork Agreggation:} In this dataset, each node is characterized by the information of other nodes connected to it (ego network). We refer to the dataset as \textbf{egoNet aggregation features} when we apply some aggregation function to the characteristics of the nodes included in the ego network. Specifically, for each attribute in the \textbf{NodeStats} feature set we apply the mean and SD in this study as in  \citet{Nargesian2017,Roa2021eswa}. Figure \ref{fig:netExample} shows how this process would be within the network; the black node corresponds to the target node, and the gray nodes belong to its Ego Network. Computing the egoNet aggregation features assumes that each connection in the ego network has the same importance; however, connections in ego networks can be weighted. 
    For this reason,  we compute the \textbf{egoNet weighted aggregation features} where the features of the neighboring nodes are weighted according to a measure of the relationship intensity measured by the weighted average and SD of the \textbf{NodeStats} attributes.
    
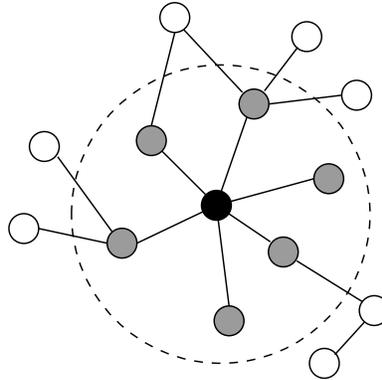
\begin{figure}[ht]
\centering
\resizebox{0.3\textwidth}{!}{%
\tikzset{every picture/.style={line width=0.75pt}} 

\begin{tikzpicture}[x=0.75pt,y=0.75pt,yscale=-1,xscale=1]

\draw  [fill={rgb, 255:red, 0; green, 0; blue, 0 }  ,fill opacity=1 ] (278.35,146.3) .. controls (278.35,140.69) and (282.89,136.15) .. (288.5,136.15) .. controls (294.11,136.15) and (298.65,140.69) .. (298.65,146.3) .. controls (298.65,151.91) and (294.11,156.45) .. (288.5,156.45) .. controls (282.89,156.45) and (278.35,151.91) .. (278.35,146.3) -- cycle ;
\draw  [fill={rgb, 255:red, 155; green, 155; blue, 155 }  ,fill opacity=1 ] (355,128.15) .. controls (355,122.54) and (359.54,118) .. (365.15,118) .. controls (370.76,118) and (375.3,122.54) .. (375.3,128.15) .. controls (375.3,133.76) and (370.76,138.3) .. (365.15,138.3) .. controls (359.54,138.3) and (355,133.76) .. (355,128.15) -- cycle ;
\draw  [fill={rgb, 255:red, 155; green, 155; blue, 155 }  ,fill opacity=1 ] (324,178.15) .. controls (324,172.54) and (328.54,168) .. (334.15,168) .. controls (339.76,168) and (344.3,172.54) .. (344.3,178.15) .. controls (344.3,183.76) and (339.76,188.3) .. (334.15,188.3) .. controls (328.54,188.3) and (324,183.76) .. (324,178.15) -- cycle ;
\draw  [fill={rgb, 255:red, 155; green, 155; blue, 155 }  ,fill opacity=1 ] (287,225.15) .. controls (287,219.54) and (291.54,215) .. (297.15,215) .. controls (302.76,215) and (307.3,219.54) .. (307.3,225.15) .. controls (307.3,230.76) and (302.76,235.3) .. (297.15,235.3) .. controls (291.54,235.3) and (287,230.76) .. (287,225.15) -- cycle ;
\draw  [fill={rgb, 255:red, 155; green, 155; blue, 155 }  ,fill opacity=1 ] (214,172.15) .. controls (214,166.54) and (218.54,162) .. (224.15,162) .. controls (229.76,162) and (234.3,166.54) .. (234.3,172.15) .. controls (234.3,177.76) and (229.76,182.3) .. (224.15,182.3) .. controls (218.54,182.3) and (214,177.76) .. (214,172.15) -- cycle ;
\draw  [fill={rgb, 255:red, 155; green, 155; blue, 155 }  ,fill opacity=1 ] (234,102.15) .. controls (234,96.54) and (238.54,92) .. (244.15,92) .. controls (249.76,92) and (254.3,96.54) .. (254.3,102.15) .. controls (254.3,107.76) and (249.76,112.3) .. (244.15,112.3) .. controls (238.54,112.3) and (234,107.76) .. (234,102.15) -- cycle ;
\draw   (250,18.15) .. controls (250,12.54) and (254.54,8) .. (260.15,8) .. controls (265.76,8) and (270.3,12.54) .. (270.3,18.15) .. controls (270.3,23.76) and (265.76,28.3) .. (260.15,28.3) .. controls (254.54,28.3) and (250,23.76) .. (250,18.15) -- cycle ;
\draw   (340,31.15) .. controls (340,25.54) and (344.54,21) .. (350.15,21) .. controls (355.76,21) and (360.3,25.54) .. (360.3,31.15) .. controls (360.3,36.76) and (355.76,41.3) .. (350.15,41.3) .. controls (344.54,41.3) and (340,36.76) .. (340,31.15) -- cycle ;
\draw   (374,71.15) .. controls (374,65.54) and (378.54,61) .. (384.15,61) .. controls (389.76,61) and (394.3,65.54) .. (394.3,71.15) .. controls (394.3,76.76) and (389.76,81.3) .. (384.15,81.3) .. controls (378.54,81.3) and (374,76.76) .. (374,71.15) -- cycle ;
\draw   (386,218.15) .. controls (386,212.54) and (390.54,208) .. (396.15,208) .. controls (401.76,208) and (406.3,212.54) .. (406.3,218.15) .. controls (406.3,223.76) and (401.76,228.3) .. (396.15,228.3) .. controls (390.54,228.3) and (386,223.76) .. (386,218.15) -- cycle ;
\draw  [dash pattern={on 4.5pt off 4.5pt}] (189.7,152.3) .. controls (189.7,96.08) and (235.28,50.5) .. (291.5,50.5) .. controls (347.72,50.5) and (393.3,96.08) .. (393.3,152.3) .. controls (393.3,208.52) and (347.72,254.1) .. (291.5,254.1) .. controls (235.28,254.1) and (189.7,208.52) .. (189.7,152.3) -- cycle ;
\draw  [fill={rgb, 255:red, 155; green, 155; blue, 155 }  ,fill opacity=1 ] (304,77.15) .. controls (304,71.54) and (308.54,67) .. (314.15,67) .. controls (319.76,67) and (324.3,71.54) .. (324.3,77.15) .. controls (324.3,82.76) and (319.76,87.3) .. (314.15,87.3) .. controls (308.54,87.3) and (304,82.76) .. (304,77.15) -- cycle ;
\draw    (251.3,109.2) -- (288.5,146.3) ;
\draw    (309.3,86.2) -- (288.5,146.3) ;
\draw    (325.3,171.2) -- (288.5,146.3) ;
\draw    (297.15,215) -- (288.5,146.3) ;
\draw    (234.3,172.15) -- (288.5,146.3) ;
\draw    (355,128.15) -- (288.5,146.3) ;
\draw    (306.3,69.2) -- (266.15,26.3) ;
\draw    (321.3,68.2) -- (344,38.4) ;
\draw    (324.3,77.15) -- (374,71.15) ;
\draw    (343.3,184.2) -- (388.3,212.2) ;
\draw    (260.15,28.3) -- (244.15,92) ;
\draw   (352,254.15) .. controls (352,248.54) and (356.54,244) .. (362.15,244) .. controls (367.76,244) and (372.3,248.54) .. (372.3,254.15) .. controls (372.3,259.76) and (367.76,264.3) .. (362.15,264.3) .. controls (356.54,264.3) and (352,259.76) .. (352,254.15) -- cycle ;
\draw   (147,162.15) .. controls (147,156.54) and (151.54,152) .. (157.15,152) .. controls (162.76,152) and (167.3,156.54) .. (167.3,162.15) .. controls (167.3,167.76) and (162.76,172.3) .. (157.15,172.3) .. controls (151.54,172.3) and (147,167.76) .. (147,162.15) -- cycle ;
\draw   (161,106.15) .. controls (161,100.54) and (165.54,96) .. (171.15,96) .. controls (176.76,96) and (181.3,100.54) .. (181.3,106.15) .. controls (181.3,111.76) and (176.76,116.3) .. (171.15,116.3) .. controls (165.54,116.3) and (161,111.76) .. (161,106.15) -- cycle ;
\draw    (167.3,162.15) -- (214,172.15) ;
\draw    (180.3,113.2) -- (217.3,164.2) ;
\draw    (369.3,248.2) -- (389.3,226.2) ;

\end{tikzpicture}}
\caption{Example of Network Features}
\label{fig:netExample}
\end{figure}    
    
\begin{figure}
\centering
\resizebox{0.5\textwidth}{!}{%
\input{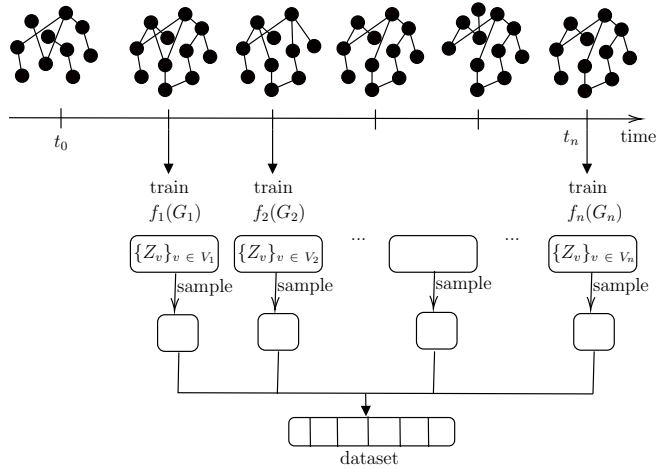}}
\caption{Node2Vec to Features}
\label{fig:N2V2Features}
\end{figure}

    \item \textbf{[N2V] Node2Vec Features:} Node2Vec  is an unsupervised method that only uses the network structure to generate the graph embedding. For the static network \textbf{FamilyNet}, Node2Vec is applied only once. A node is characterized by this embedding regardless of the moment it was sampled in the dataset.
    For temporal networks (\textbf{EOWNet}), Node2Vec has to be recomputed every period because of the network changes. Each node is characterized by the embedding corresponding to the month in which it was sampled in the dataset. Figure \ref{fig:N2V2Features} shows the process through which to obtain the embedding features by applying Node2Vec.  Each period, a new model is trained, and the resulting embedding is consolidated to characterize the sample dataset that will allow us to train the final credit scoring models.

\begin{figure}
\centering
\resizebox{0.5\textwidth}{!}{%
\input{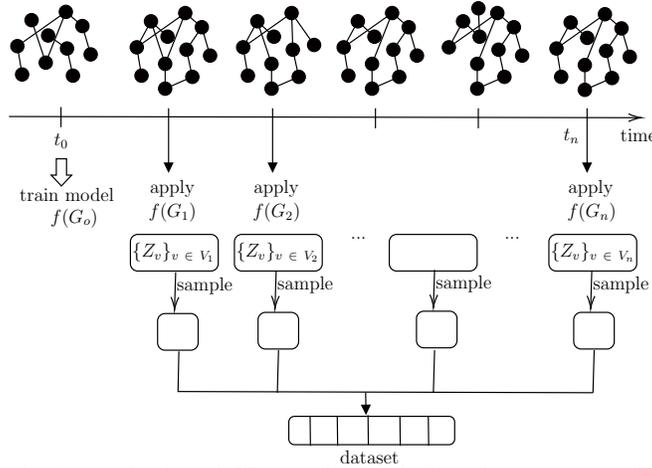}}
\caption{Graph Convolutional Networks and Graph Autoencoders to Features}
\label{fig:GNC2Features}
\end{figure}

    \item \textbf{[GNN] Graph Neural Network Features:} We used either GCN or GAE for the extraction of GNN features. These methods carry out the graph convolution of the network structure through a node feature vector in order to generate as  output the feature vector for ensuing classification by machine learning approaches. In this study, GNN input feature vectors are in the \textbf{Node} feature dataset. Regardless of whether the network is static or temporal, the feature vector is dynamic and changes during each observation period, that is, each month.
     To avoid target leakage \citep{Kaufman2011}, we train the GCNs  with the first available data period; these models are  applied in the subsequent months while the network or the feature vectors change.  This approach does not handle new entrants to the social interaction networks; however, due to the sources of  social interaction information employed in this study, most  thin-file borrowers are taken into account in the training dataset, despite having temporal networks. Therefore, new entrants do not affect our results. Under other circumstances, for instance when working with a partial network, our recommendation is to train a new model for every period, taking precautions not to incur in target leakage on the population of the credit scoring model. Depending on the dynamism of the network, one solution is to calculate local connection updates as suggested in \citet{Vlasselaer2015} in the interim between model training phases.
    The GCNs are trained on the entire network (either FamilyNet or EOWNet) regardless of whether the nodes belong to our training dataset or not. The output of the GCN is the label of the nodes in the network, which can be either  defaulter, non-defaulter, or unbanked; therefore, each GCN solves a multi-classification problem by providing the a posteriori probabilities of each label for each node. Finally, each node is characterized by the GCN Features resulting from the application of the GCN on the network and on the feature vector of the month in which the node was sampled, that is, in which it entered the banking system. Figure \ref{fig:GNC2Features} illustrates the feature-engineering process to extract the GCN features.
    Regarding GAE, the feature engineering process is similar. In this case, the embedding corresponds to the bottleneck hidden layer representing the encoder section's output. For consistency, we  apply the same data selection applied for the GCN to the training of the GAE models, although their training is unsupervised.

\end{itemize}

\subsection{Feature Subsets} \label{sec:RfeatSetup}

The datasets for machine learning training and validation experiments are composed of the following subsets of features. 
\begin{itemize}
    \item Subset $\textbf{A}: X_{Node}$ is the dataset of node characteristics.
    \item Subset $\textbf{B}: X_{BenchScore}$ is the benchmark score; this attribute is also used as a benchmark to quantify the performance of our proposed approach.
    \item Subset $\textbf{C}: X_{NodeStats}$ is composed of the statistics obtained from the position of the node within the network.
    \item Subset $\textbf{D}: X_{EgoNet}$ includes the egoNet aggregation and the egoNet weighted aggregation features that are calculated in three scenarios, considering the entire network, considering only those edges that are bridges, and considering those edges that are not bridges\footnote{An edge connecting the nodes $u$ and $v$ is called a \textbf{bridge}; if removing this edge, there is no longer a path connecting $u$ and $v$}.
    \item Subset $\textbf{E}: X_{GNN, N2V}$ corresponds to the features created by applying GNNs and Node2Vec. People are characterized by features from both networks (EOWNet and FamilyNet), while companies are characterized only by EOWNet.
\end{itemize}

For this study, we aggregated these feature subsets into eight increasingly larger datasets for training and validation, each defining a different experimental setting. The details of the feature sets are presented below in Table \ref{table:featuresGroup}.

\begin{table}[ht]
\footnotesize
\centering
\caption{Experiments Setup}
\label{table:featuresGroup}
\begin{tabular}{ll}
\toprule
   Experiment Id  &    Feature Group\\ \midrule
     
    A & $ X = \{ X_{Node} \}$ \\
    A+B &$ X = \{ X_{Node} + X_{BenchScore}\}$ \\
    
    A+B+C & $ X = \{ X_{Node} + X_{BenchScore} + X_{NodeStats} \}$ \\
    A+B+D & $ X = \{ X_{Node} + X_{BenchScore} + X_{EgoNet} \}$ \\
    A+B+E & $ X = \{ X_{Node} + X_{BenchScore} +  X_{GNN,N2V} \}$    \\
   A+B+C+D & $ X = \{ X_{Node} + X_{BenchScore} + X_{NodeStats} + X_{EgoNet} \}$ \\
   A+B+C+E & $ X = \{ X_{Node} + X_{BenchScore} + X_{NodeStats} + X_{GNN,N2V} \}$ \\
   A+B+C+D+E & $ X = \{ X_{Node} + X_{BenchScore} + X_{NodeStats} + X_{EgoNet} + X_{GNN,N2V} \}$ \\ \bottomrule    
\end{tabular}
\end{table}

\subsection{Evaluation Metrics}\label{sec:evaluationmetrics}

The AUC is a  popular metric used to evaluate the model performance in classification problems \citep{Zeng2019}.  It ranges between 0.5 and 1. Values closer to 1 indicate a better discriminatory capacity, while a value of 0.5 indicates a performance equivalent to a chance decision. In the context of credit scoring, the AUC can be easily interpreted as follows:  For a randomly selected defaulter and  non-defaulter pair, the AUC corresponds to the probability that the classification model assigns a higher score to the defaulter.

Another extensively utilized performance measure is the Kolmogorov-Smirnov (KS) statistic, which measures the distance separating the cumulative distributions of defaulters $(P_{D}(t))$ and non-defaulters $(P_{ND}(t))$ \citep{Fang2019}. The KS statistic is defined as:
\begin{equation}
    KS = \max_{t}|P_{D}(t) - P_{ND}(t)|
\end{equation}

and KS ranges between 0 and 1, and  a higher KS indicates a higher prediction performance. 

\subsection{Methodology}\label{sec:Method}

First, we carry out a \textit{feature engineering} process that seeks to create attributes to characterize the nodes from the network. Then, in \textit{the train-test split} step, the available dataset is divided into a   training dataset of which 30\% consists of the samples used to estimate the model's hyper-parameters, and the remaining   70\% consists of the samples used to train and validate models according to an N-Fold Cross-Validation scheme.

Before estimating the best hyper-parameters, we apply a feature selection process. In this step, the intention is to choose a low-correlated subset of features with high predictive power. Three selection levels are formulated. 
The first is a bivariate selection that only considers one  feature at a time and the target vector to build a prediction model, evaluating its performance. The selection process applies a threshold to this feature's predictive power. Only those variables are selected such that $KS > KS_{min}$ and $ AUC > AUC_{min}$, where $KS_{min}$ and  $AUC_{min}$ are threshold parameters. 

Next, a multivariate selection is   applied following   a simple but effective method to drop correlated features that we have developed. This algorithm starts with an empty list $\mathcal{S}$. We iterate over a set of features $\mathcal{P}$ in decreasing order of predictive power and append to $\mathcal{S}$ those features whose absolute value of the correlation with each and all the features in $\mathcal{S}$  is less than a threshold  $\rho$  set to avoid high correlated features \citep{Akoglu2018}. The algorithm stops when all the features have been visited. The first feature in $\mathcal{P}$ is added to $\mathcal{S}$ without correlation comparison.

For this study, the multivariate selection process is applied twice. In the  first application, the process selects low-correlated features for each group of attributes $\mathcal{P}\in \{  X_{Node} \cup  X_{BenchScore},  \allowbreak   X_{NodeStats},\allowbreak  X_{EgoNet}, \allowbreak  X_{GNN,N2V} \}.   $ Secondly, it is applied globally to all the remaining features $\mathcal{P} = \{X_{Node} \cup  X_{BenchScore}  \cup X_{NodeStats} \allowbreak  \cup  X_{EgoNet} \cup X_{GNN,N2V}\}  $. In both cases, a threshold $\rho $ is used, and the features are ordered by the features' AUC, from higher to lower.

Finally, at the N-Fold Cross-Validation stage, the dataset is partitioned into N subsets of equal size. Each subset is used alternatively as the test dataset, while the remaining folds are used to train the classification model. The hyper-parameters used in each iteration are  those estimated in the previous stage. Additionally, in each of these iterations, multiple models are trained with different feature sets and stored to be used later to compare the models.

\subsection{Experimental Setup}
The parameters of the univariate selection are set at $ KS_ {min} = 0.01 $ and $ AUC_ {min} = 0.53 $; for the multivariate selection process, $ \rho = 0.7 $ in both processes to avoid high correlated features \citep{Akoglu2018}. The N-Fold Cross-Validation stage is carried out considering $ N = 10 $, and in each iteration, the results of gradient boosting \citep{Friedman2001} models are displayed. Other classification models such as regularized logistic regression and Random Forest \citep{Breiman2001} were trained. However, gradient boosting consistently delivered better results.

\section{Results and Discussion}\label{sec:Results}
In this section, we present the results obtained. We begin with the technical implementation details; then, we analyze the execution times. Subsequently, we display the model's performance in three scenarios: the impact on performance using traditional features, the impact on performance using the different graph representation methods, and the advantages of combining these methods. Finally, an analysis is presented of the main features, traditional and network-based, for the creditworthiness assessment.

\subsection{Implementation Details}\label{sec:ImpDetails}

In this work, we used the Python implementations  Networkx v2.6.3 \citep{Hagberg2008} and Stanford Network Analysis Platform (SNAP) v5.0.0 \citep{Leskovec2016snap}  in the hand-crafted feature engineering process ($ X_{NodeStats},\;  \allowbreak  X_{EgoNet}$); for Node2Vec, GCN and GAE ($ X_{GNN,N2V}$) we used PyTorch v1.6.0 \citep{Paszke2019} and PyTorch Geometric v2.0.1 \citep{Fey2019}. 

 To conduct the experiments, we used a laptop with an Intel 8-Core i7 CPU and 32 GB of RAM for network construction and hand-crafted feature engineering. For the Node2Vec, GCN, GAE, and model training phases, we used a server with a driver node with 140 GB of RAM and 20 CPU cores and between two and eight auto-scaling worker nodes with 112GB of RAM and 16 CPU cores.

\subsection{Execution Time}
Below we detail the execution time of the most critical stages of our work, the implementation of the GRL methods, and the models' training.

\begin{itemize}
    \item \textbf{[NodeStats] Node Statistics:} This process corresponds to the computation of the metrics defined in Section \ref{sec:featEng}. This process is carried out only once for the static network FamilyNet, and it is calculated for all the available periods (24) of the EOWNet. The computation of all the metrics for a network takes, on average, 25 minutes. The total execution time of this stage was 625 minutes.
    
    \item \textbf{[EgoNet] EgoNetwork Aggregation:} This process is calculated once per network type (FamilyNet and EOWNet). The total execution time of this stage was 300 minutes.
    
    \item \textbf{[N2V] Node2Vec Features:} This process is carried out only once for the static network FamilyNet, and it is calculated for all the available periods (24) of the EOWNet. The Node2Vec training for a network takes, on average, 300 minutes. The total execution time of this stage was 7,500 minutes.
    
    \item \textbf{[GNN] Graph Neural Network Features:} In this stage, eight models are trained using each network and eight different feature vectors. The training of each model  is carried out over the first available period data; afterward, the trained model is applied to the data of the remaining periods. Table \ref{table:GNC_GAE_ET} shows the execution time by GNN type and by network type. The total execution time of this stage was 6,920 minutes.

\begin{table}[ht]
\footnotesize
\centering
\caption{GCN and GAE Execution Time}
\label{table:GNC_GAE_ET}
\resizebox{0.75\textwidth}{!}{%
\begin{tabular}{|c|c|r|r|r|}
\hline
GNN Type&Network Type& \parbox{1.8cm}{\centering Unit Train Time (min)}&\parbox{1.8cm}{\centering Total Training Time (min)}& \parbox{1.8cm}{\centering Total Apply Time (min)}\\ \hline
\multirow{2}{*}{{\centering GCN}}&EOWNet&25&200&600 \\ 
&FamilyNet&60&480&600 \\ \hline
\multirow{2}{*}{{\centering GAE}}&EOWNet&140&1,120&1,200 \\ 
&FamilyNet&190&1.520&1.200 \\ \hline
\multicolumn{3}{|c|}{Total Execution Time (min)}&3,320&3,600 \\
\hline
\end{tabular}}
\end{table}

    \item \textbf{Gradient boosting training:} Four models were trained using the methodology described in Section \ref{sec:Method} for the scenarios defined in Section \ref{sec:datadescr}, predicting application and behavioral scoring for individuals and companies. The complete training for each scenario takes, on average, 40 minutes. The total execution time of this step was 160 minutes.

\end{itemize}

The total execution time of the application of the proposed  methodology to the datasets is 15,500 minutes. Although a large part of these executions was parallelized using the server described in Section \ref{sec:ImpDetails}, the high computational cost is mainly due to two factors, the volume of data and the complexity of the algorithms. Regarding the large volume of data, the FamilyNet has 20 million nodes and 30 million edges, and the EOWNet has 8.6 million nodes and 26 million edges. These massive dataset sizes directly impact the algorithms used, because the complexity depends on the nodes $| V |$, edges $| E |$ and the embedding dimension $d$; the algorithmic complexity for this particular case is the following: Node2Vec: $ \mathcal {O} (| V | d) $ \citep {Grover2016}, Graph Convolutional Networks: $ \mathcal {O} (| E | d) $ \citep {Kipf2016} and Graph Autoencoder: $ \mathcal {O} (| V | ^ 2 d) $ \citep {Kipf2016VGAE}.

\subsection{Model Performance Results}

The model training process was conducted for each Experiment ID using different Feature Set combinations as specified in Section \ref{sec:Rdataset}. The results are presented in Tables \ref{table:resultsExpAUC} and \ref{table:resultsExpKS}. These tables show the relative improvement in AUC and KS achieved by each model over  the baseline BenchScore, measured as $\frac{row_{AUC}-BenchScore_{AUC}}{BenchScore_{AUC}}$ and $\frac{row_{KS}-BenchScore_{KS}}{BenchScore_{KS}} $, respectively. Each row corresponds to an experiment defined in Section \ref{sec:RfeatSetup}, and each column displays the four credit scoring scenarios illustrated in Section \ref{sec:Rdataset}.

 The reported results correspond to the average of 10-fold cross-validation, as indicated in Section \ref{sec:ExpDesign}. A t-test is applied to establish the statistical significance of the performance differences obtained using the different feature sets, according to the suggestions proposed by \citet{Flach2012}.

\begin{table}[ht]
\footnotesize
\centering
\caption{Improvement in AUC relative to the benchmark model (mean and std). We only report results when the equal performance hypothesis is rejected, with a confidence level of 95\%; otherwise, we display *. The best performance in each column is shown in bold; more than one bold value indicates that the hypothesis of equal performance between those models cannot be rejected.}
\label{table:resultsExpAUC}
\resizebox{0.70\textwidth}{!}{%
\begin{tabular}{|l|r|r|r|r|}
\hline
\multirow{2}{*}{\parbox{1.5cm}{\centering Feature Set}} &     \multicolumn{2}{c|}{Business Credit Score} &    \multicolumn{2}{c|}{Personal Credit Score}\\ \cline{2-5}
     &      Application &    Behavioral &      Application &    Behavioral \\ \hline
          A & -3.52\%          $\pm$ 2.87\%& -0.90\% $\pm$ 0.21\% & -0.74\% $\pm$ 0.63\% & -0.63\% $\pm$ 0.09\% \\ \hline
        A+B &  *                     &  0.58\% $\pm$ 0.06\% &  1.45\% $\pm$ 0.39\% &  0.95\% $\pm$ 0.06\% \\ \hline
      A+B+C &  *                     &  1.13\% $\pm$ 0.12\% &  2.02\% $\pm$ 0.49\% &  1.07\% $\pm$ 0.06\% \\ \hline
      A+B+D &  \textbf{8.96}\% $\pm$ 3.37\% &  2.33\% $\pm$ 0.15\% &  2.31\% $\pm$ 0.64\% &  1.25\% $\pm$ 0.08\% \\ \hline
      A+B+E &  3.92\%          $\pm$ 2.03\% &  1.77\% $\pm$ 0.13\% &  3.17\% $\pm$ 0.55\% &  1.96\% $\pm$ 0.04\% \\ \hline
    A+B+C+D &  \textbf{9.00}\% $\pm$ 3.47\% &  2.37\% $\pm$ 0.16\% &  2.39\% $\pm$ 0.60\% &  1.32\% $\pm$ 0.08\% \\ \hline
    A+B+C+E &  4.25\%          $\pm$ 1.84\% &  1.94\% $\pm$ 0.16\% &  3.26\% $\pm$ 0.48\% &  2.03\% $\pm$ 0.05\% \\ \hline
  A+B+C+D+E &  \textbf{8.43}\% $\pm$ 2.83\% & \textbf{2.80}\% $\pm$ 0.16\% &   \textbf{3.58}\% $\pm$ 0.61\% &  \textbf{2.18}\% $\pm$ 0.04\%\\
  \hline
\end{tabular}}
\end{table}

\begin{table}[ht]
\footnotesize
\centering
\caption{Improvement in KS relative to the benchmark model (mean and std). We only report results when the equal performance hypothesis is rejected, with a confidence level of 95\%; otherwise, we display *. The best performance in each column is shown in bold; more than one bold value indicates that the hypothesis of equal performance between those models cannot be rejected.}
\label{table:resultsExpKS}
\resizebox{0.70\textwidth}{!}{%
\begin{tabular}{|l|r|r|r|r|}
\hline
\multirow{2}{*}{\parbox{1.5cm}{\centering Feature Set}} &     \multicolumn{2}{c|}{Business Credit Score} &    \multicolumn{2}{c|}{Personal Credit Score}\\ \cline{2-5}
     &      Application &    Behavioral &      Application &    Behavioral \\ \hline
           A &  *                     & -4.15\%  $\pm$ 0.94\% &  -5.25\%  $\pm$ 2.40\% & -2.39\%  $\pm$ 0.46\% \\ \hline
        A+B &  *                      &  1.56\%  $\pm$ 0.40\% &   4.38\%  $\pm$ 1.19\% &  1.95\%  $\pm$ 0.35\% \\ \hline
      A+B+C &  *                      &  3.21\%  $\pm$ 0.71\% &   6.27\%  $\pm$ 1.02\% &  2.23\%  $\pm$ 0.39\% \\ \hline
      A+B+D &  \textbf{20.69}\%  $\pm$ 16.73\% &  7.69\%  $\pm$ 0.92\% &   6.79\%  $\pm$ 1.36\% &  2.69\%  $\pm$ 0.47\% \\ \hline
      A+B+E &  \textbf{12.22}\%  $\pm$ 10.89\% &  5.83\%  $\pm$ 0.74\% &   8.64\%  $\pm$ 2.13\% &  4.68\%  $\pm$ 0.28\% \\ \hline
    A+B+C+D &  \textbf{21.28}\%  $\pm$ 17.10\% &  8.09\%  $\pm$ 0.95\% &   7.12\%  $\pm$ 1.52\% &  2.83\%  $\pm$ 0.52\% \\ \hline
    A+B+C+E &  \textbf{12.88}\%  $\pm$ 10.11\% &  6.33\%  $\pm$ 0.70\% &   8.93\%  $\pm$ 1.98\% &  4.93\%  $\pm$ 0.26\% \\ \hline
  A+B+C+D+E &  \textbf{19.32}\%  $\pm$  14.77\% &  \textbf{9.45}\% $\pm$ 0.85\% &  \textbf{10.83}\%  $\pm$ 1.98\% &  \textbf{5.15}\% $\pm$ 0.42\% \\  
  \hline
\end{tabular}}
\end{table}

\subsubsection{Model Performance Using Traditional Features}

The first step toward reaching a conclusion on the contribution of network data is to understand whether our methodology enables us to obtain equal or better results than the current decision-making scheme in the financial institution. For this, we compare the BenchScore with the results obtained by the feature set \textbf{A+B}. A comparison with only the feature set \textbf{A} is not entirely accurate, considering that we do not have access to all the features used in the BenchScore training.

The results show that our methodology obtains equal or greater performance, measured in terms of AUC and KS statistics, in all four scenarios; three of them are greater, with statistically significant differences.

The performance enhancements in Behavioral Business Credit Scoring are 0.58\% and 1.56\% for AUC and KS, respectively. In terms of Personal Credit Scoring, the performance enhancements are AUC: 1.45\%, KS: 4.38\% for Application Scoring and AUC: 0.95\%, KS: 1.95\% for Behavioral Scoring. 
These results indicate that using our methodology and training a model with similar features to the benchmark model can obtain better results than the current decision scheme applied by the financial institution.

\subsubsection{Model Performance Using Graph Representation Learning Features}

In Application Business Credit Scoring, a nearly 9\% AUC increase over the BenchScore is achieved. These results are obtained by a model that incorporates three feature sets: A+B+D, A+B+C+D, and A+B+C+D+E. Note that in these three sets, the common attributes correspond to the traditional features $\textbf{A+B}: X_{Node} \cup X_{BenchScore}$  and EgoNet Aggregation Features $\textbf{D}: X_{EgoNet}$. The performance comparison between all feature sets is shown in Table \ref{table:PJ_NEW_ROC}; we marked as * those comparisons with no statistically significant differences.

\begin{table}[ht]
\footnotesize
\centering
\caption{Performance comparison of Business Application Scoring. Performance is measured by the relative increase in AUC $(\frac{row_{AUC} - column_{AUC}}{column_{AUC}})$.
}
\label{table:PJ_NEW_ROC}
\resizebox{0.85\textwidth}{!}{%
\begin{tabular}{lrrrrrrrrr}
\toprule
& BENCH & A & A+B & A+B+C & A+B+D & A+B+E & A+B+C+D & A+B+C+E & A+B+C+D+E\\
\midrule
BENCH &  *&  3.65\%&  *&  *&  -8.23\%&  -3.77\%&  -8.26\%&  -4.08\%&  -7.77\%\\ 
A &  -3.52\%&  *&  -2.92\%&  -4.55\%&  -11.45\%&  -7.16\%&  -11.49\%&  -7.45\%&  -11.01\%\\ 
A+B &  *&  3.01\%&  *&  -1.68\%&  -8.79\%&  -4.37\%&  -8.83\%&  -4.67\%&  -8.34\%\\ 
A+B+C &  *&  4.77\%&  1.71\%&  *&  -7.23\%&  -2.73\%&  -7.26\%&  -3.04\%&  -6.77\%\\ 
A+B+D &  8.96\%&  12.94\%&  9.64\%&  7.79\%&  *&  4.85\%&  *&  4.52\%&  *\\ 
A+B+E &  3.92\%&  7.71\%&  4.57\%&  2.81\%&  -4.63\%&  *&  -4.66\%&  *&  -4.15\%\\ 
A+B+C+D &  9.00\%&  12.98\%&  9.68\%&  7.83\%&  &  4.89\%&  *&  4.56\%&  *\\ 
A+B+C+E &  4.25\%&  8.05\%&  4.90\%&  3.13\%&  -4.32\%&  *&  -4.36\%&  *&  -3.85\%\\ 
A+B+C+D+E &  8.43\%&  12.38\%&  9.10\%&  7.26\%&  *&  4.33\%&  *&  4.00\%&  *\\ 
\bottomrule
\end{tabular}}
\end{table}

When performance is measured in terms of KS, the maximum is obtained with five feature sets, with GRL features, but no method for feature extraction predominates. Although differences are observed in the values presented, these are not statistically significant.
From these results, it is necessary to highlight at least one GRL method in the best feature sets. The complete comparison is presented in Table \ref{table:PJ_NEW_KS}

\begin{table}[ht]
\footnotesize
\centering
\caption{Performance comparison of Business Application Scoring. Performance is measured by the relative increase in KS $(\frac{row_{KS} - column_{KS}}{column_{KS}})$.}
\label{table:PJ_NEW_KS}
\resizebox{0.9\textwidth}{!}{%
\begin{tabular}{lrrrrrrrrr}
\toprule
& BENCH & A & A+B & A+B+C & A+B+D & A+B+E & A+B+C+D & A+B+C+E & A+B+C+D+E\\ \midrule
BENCH &  *&  *&  *&  *&  -17.14\%&  -10.89\%&  -17.55\%&  -11.41\%&  -16.19\%\\ 
A &  *&  *&  *&  *&  -19.94\%&  -13.90\%&  -20.34\%&  -14.40\%&  -19.02\%\\ 
A+B &  *&  *&  *&  *&  -16.16\%&  -9.83\%&  -16.57\%&  -10.36\%&  -15.20\%\\ 
A+B+C &  *&  *&  *&  *&  -16.02\%&  -9.68\%&  -16.43\%&  -10.21\%&  -15.06\%\\ 
A+B+D &  20.69\%&  24.91\%&  19.28\%&  19.08\%&  *&  *&  *&  *&  *\\ 
A+B+E &  12.22\%&  16.14\%&  10.90\%&  10.72\%&  *&  *&  *&  *&  *\\ 
A+B+C+D &  21.28\%&  25.53\%&  19.86\%&  19.67\%&  *&  *&  *&  *&  *\\ 
A+B+C+E &  12.88\%&  16.83\%&  11.56\%&  11.37\%&  *&  *&  *&  *&  *\\ 
A+B+C+D+E &  19.32\%&  23.49\%&  17.92\%&  17.73\%&  *&  *&  *&  *&  *\\ 
\bottomrule
\end{tabular}}
\end{table}

The best performance is observed in other scenarios when combining traditional features and all the GRL features; this corresponds to the Feature Set \textbf{A+B+C+D+E}. The best performance is achieved in AUC (see Table \ref{table:resultsExpAUC}) and KS (see Table \ref{table:resultsExpKS}).

These results are of great importance because they indicate that the methods combined by our methodology are complementary, and none is significantly better than   the others. Both methods, namely hand-crafted feature engineering and GNNs, have until now been treated in the literature as independent in addressing the credit scoring problem.

When comparing the results of Application and Behavioral Credit Scoring, it is observed that the most significant increase in performance, regardless of the metric, is achieved in Application Credit Scoring. Network-related features complement the least availability of information, such that the relationships that a person or company has are relevant when predicting their creditworthiness. These results are of high interest for lenders and in terms of their strategies for the unbanked. The improvement in predictive performance implies that   more borrowers can be serviced without increasing the portfolio default rate.

Regarding Behavioral Credit Scoring, traditional attributes are already good predictors of creditworthiness; the borrower's credit behavior is a good indicator of   default. For this reason, the increase in predictive performance is more limited, although still significant.

\subsubsection{The Advantages of Blending Graph Representation Learning}

The previous sections have shown that our approach allows us to enhance the discrimination power of our benchmark in terms of AUC and KS. Through the incorporation of the graph data by means of the GRL methods, this increase is even more significant. Now, we are interested in discovering the contribution of each of these methods. The performance comparison between the A+B+C+D+E feature set and each method by itself is shown in Tables \ref{table:BLEND_AUC} and \ref{table:BLEND_KS}, for AUC and KS respectively; we marked as * those comparisons with no statistically significant differences. In each table, the results are presented for each credit scoring scenario and the comparison using the $X_{EgoNet}$ (A + B + D) and $X_{GNN,N2V}$ (A + B + E) features; In both cases, the models trained with the $X_{NodeStats}$ features are also included.

\begin{table}[ht]
\footnotesize
\centering
\caption{Blended Graph Representation Learning performance. The performance enhancement of training a model using all GRL methods (A+B+C+D+E) is measured as the relative increase in AUC given by $(\frac{[A+B+C+D+E]_{AUC} - column_{AUC}}{column_{AUC}})$.}
\label{table:BLEND_AUC}
\resizebox{0.9\textwidth}{!}{%
\begin{tabular}{|r|c|rr|rr|}
\hline
\multirow{2}{*}{\parbox{2.5cm}{\centering Scoring}} &\multirow{2}{*}{\parbox{1.5cm}{\centering Model}} &\multicolumn{4}{c|}{Feature Set}\\ \cline{3-6}
 &  &A+B+D  &A+B+C+D    &A+B+E  &A+B+C+E    \\ \hline
\multirow{2}{*}{\parbox{2.5cm}{\centering Application Scoring}} &Business Credit Score   &*       &*      &4.33\%     &4.00\% \\ \cline{2-6}
                                                                            &Personal Credit Score   &1.23\%  &1.16\% &0.39\%     &0.31\%\\ \hline
\multirow{2}{*}{\parbox{2.5cm}{\centering Behavioral Scoring}}              &Business Credit Score   &0.47\%  &0.43\% &1.02\%     &0.85\%\\\cline{2-6}
                                                                            &Personal Credit Score   &0.92\%  &0.84\% &0.22\%     &0.15\%\\
\hline
\end{tabular}}
\end{table}

\begin{table}[ht]
\footnotesize
\centering
\caption{Blended Graph Representation Learning performance. The performance enhancement of training a model using all GRL methods (A+B+C+D+E) is measured as the relative increase in KS $(\frac{[A+B+C+D+E]_{KS} - column_{KS}}{column_{KS}})$.}
\label{table:BLEND_KS}
\resizebox{0.9\textwidth}{!}{%
\begin{tabular}{|r|c|rr|rr|}
\hline
\multirow{2}{*}{\parbox{2.5cm}{\centering Scoring}} &\multirow{2}{*}{\parbox{1.5cm}{\centering Model}} &\multicolumn{4}{c|}{Feature Set}\\ \cline{3-6}
 &  &A+B+D  &A+B+C+D    &A+B+E  &A+B+C+E    \\ \hline
\multirow{2}{*}{\parbox{2.5cm}{\centering Application Scoring}} &Business Credit Score   &*       &*          &*             &* \\ \cline{2-6}
                                                                            &Personal Credit Score   &3.79\%  &3.47\% &2.02\%     &1.75\%\\ \hline
\multirow{2}{*}{\parbox{2.5cm}{\centering Behavioral Scoring}}              &Business Credit Score   &1.68\%  &1.31\% &3.47\%     &2.99\%\\\cline{2-6}
                                                                            &Personal Credit Score   &2.40\%  &2.26\% &0.45\%     &0.21\%\\
\hline
\end{tabular}}
\end{table}

The results show that combining the GRL methods always generates better or similar results than using each method independently. An equal performance is only obtained for the Business Application Credit Score, where the only statistically significant increase, in AUC terms, occurs when using the $ X_{GNN,N2V}$ features. However, this feature subset does not produce an increment compared to using only the $X_{EgoNet}$ features. 
On the other hand, in all other scenarios, the GRL combination generates statistically significant increments, independent of the method used and whether or not the  $X_{NodeStats}$ features are incorporated.
In this way, our approach allows us to increase discriminatory power in assessing creditworthiness, generate more accurate models, and use graph data better through a framework that combines multiple methods of GRL.

\subsection{Feature Importance Analysis}

To determine the importance of each feature, we utilize SHAP: SHapley Additive exPlanations \citep{lundberg2017}, an approach based on game theory that calculates each attribute's importance by comparing the model predictions with and without the attribute. The global feature importance is examined with regard to the four scenarios described earlier, that is, the prediction of Application or Behavioral Scoring for individuals or for businesses. All analyses are conducted with the feature set \textbf{A+B+C+D+E}, which incorporates all the features and is the one that reports the best results.

\subsubsection{Business Credit Scoring }

\begin{figure}[ht]
  \centering
  \noindent
  \caption{Business Credit Scoring:  Feature Importance }\label{fig:PJ_SHAP01}
  \begin{tabular}{cc}
    \subfigure[Application: Average impact on model output]{\includegraphics[width=0.45\textwidth]{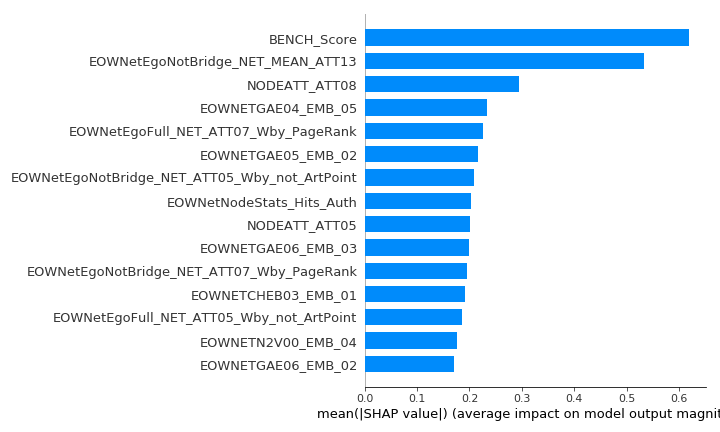}}&
    \subfigure[Application: Impact on model output]{\includegraphics[width=0.45\textwidth]{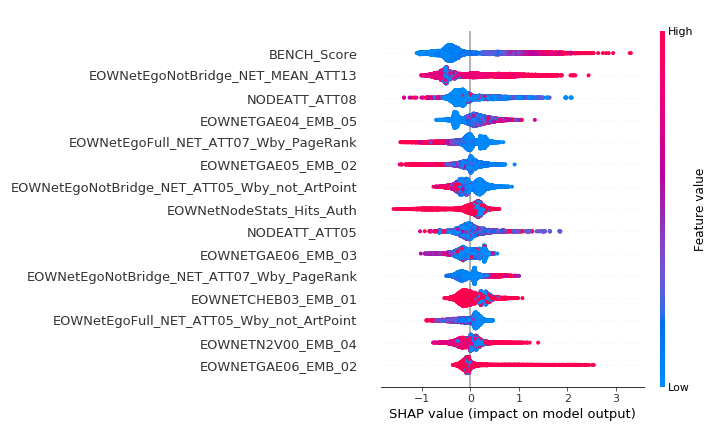}} 
    \\
    \subfigure[Behavioral: Average impact on model output]{\includegraphics[width=0.45\textwidth]{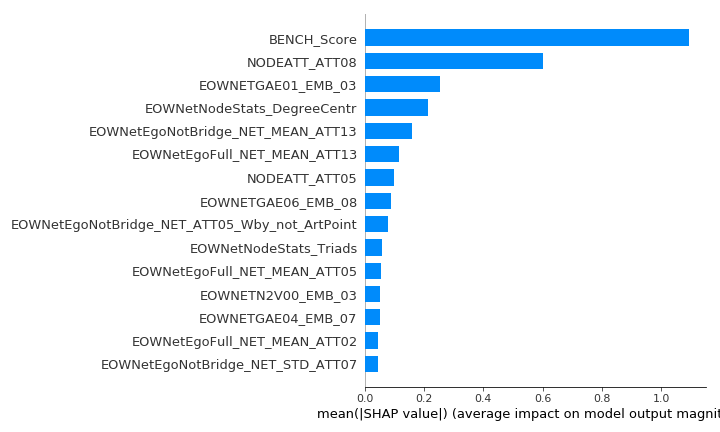}}&
    \subfigure[Behavioral: Impact on model output]{\includegraphics[width=0.45\textwidth]{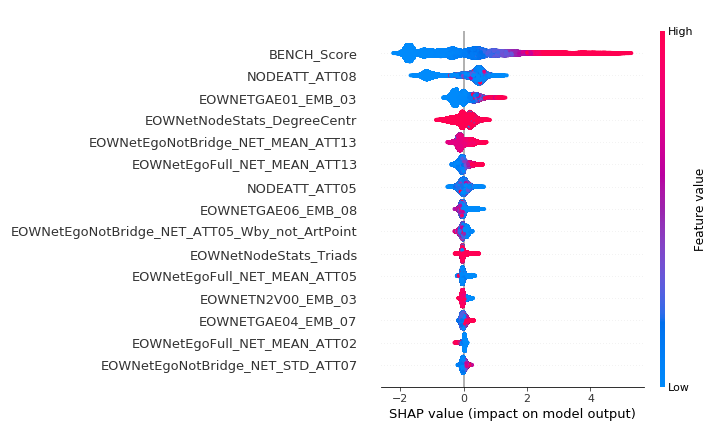}}

  \end{tabular}
\end{figure}

In Figure \ref{fig:PJ_SHAP01}, the importance of the attributes incorporated into the model is presented. Figures \ref{fig:PJ_SHAP01}(a) and \ref{fig:PJ_SHAP01}(c) show each attribute's importance for Application and Behavioral Scoring, respectively; we define it as the average of the absolute values of the SHAP values. Figures \ref{fig:PJ_SHAP01}(b) and \ref{fig:PJ_SHAP01}(d) show each feature's impact on the model output; only the 15 most relevant attributes are displayed in both figures. Figures \ref{fig:PJ_SHAP01}(b) and \ref{fig:PJ_SHAP01}(d) allow us to understand how the value of a particular feature affects the probability of default.

As expected, the most significant contribution to the model is the BenchScore, which already summarizes valuable information about each company that allows the estimation of its creditworthiness. This influence occurs in both scenarios, Application, and Behavioral. However, its importance is more significant in Behavioral Scoring.

Among these 15 relevant attributes in both scenarios, only the BenchScore, commercial debt amount (NODEATT\_08), and unused revolving credit amount  (NODEATT\_05) correspond to the business-related characteristics. The remaining top features correspond to Network-related features.

An additional relevant feature is the average BenchScore of the company's ego network, including only the non-bridge edges. This result indicates the creditworthiness of the company's neighborhood is also highly predictive of the company's creditworthiness. In Application Scoring, this feature is practically as relevant as the BenchScore. See Table \ref{table:FeatureNomenclature} for more detail on the description of the most relevant variables. This table presents the taxonomy of the features used in the current study, giving the necessary specifications for the correct interpretation of the feature attributes and the nomenclature used for the management of the datasets.

Further, we find attributes whose influence corresponds to people related to the company, including its owners, for instance, the attribute generated from a Graph Autoencoder trained with the consumer debt of the EOWNet Network. The consumer-debt effect of the ego network is also observed in the attribute corresponding to the consumer debt weighted by the PageRank of the node's neighborhood. The presence of consumer debt as a relevant network-related feature in Business Scoring is highly significant, especially in SMEs. The EgoNet's short-term personal debt, mainly on the part of the owners, accounts for the often blurred separation between personal finances and company finances.
The owner's default can affect the company and vice versa. This hypothesis requires a more detailed investigation and will be addressed for future work.

\begin{figure}[ht]
  \centering
  \noindent
  \caption{Business Credit Scoring: Treemap of Feature Importance, the Average Impact on Model Output}\label{fig:PJ_SHAP02}
  \begin{tabular}{c}
    \subfigure[Application Scoring]{\includegraphics[width=0.85\textwidth]{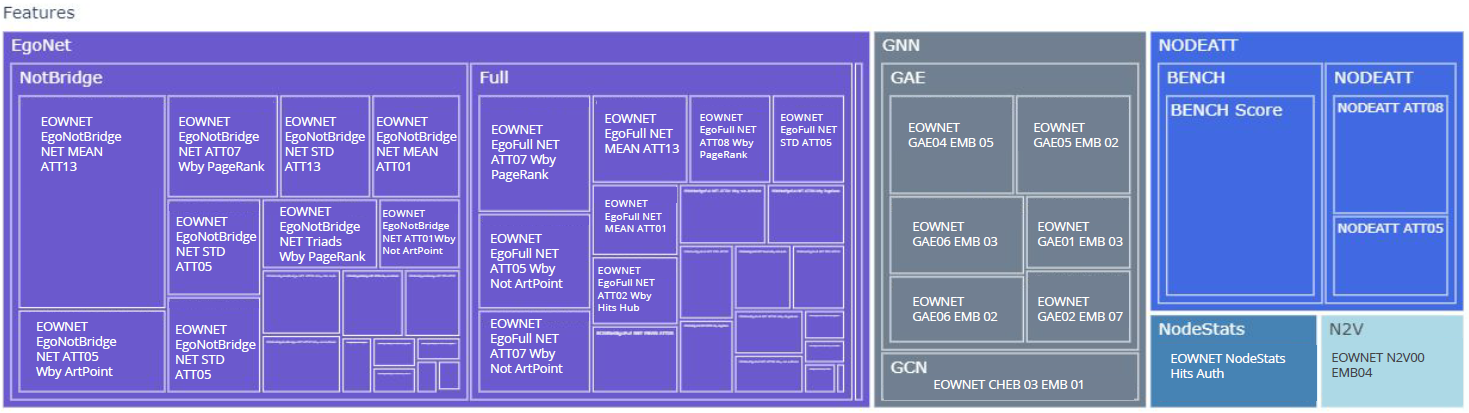}}\\
    \subfigure[Behavioral Scoring]{\includegraphics[width=0.85\textwidth]{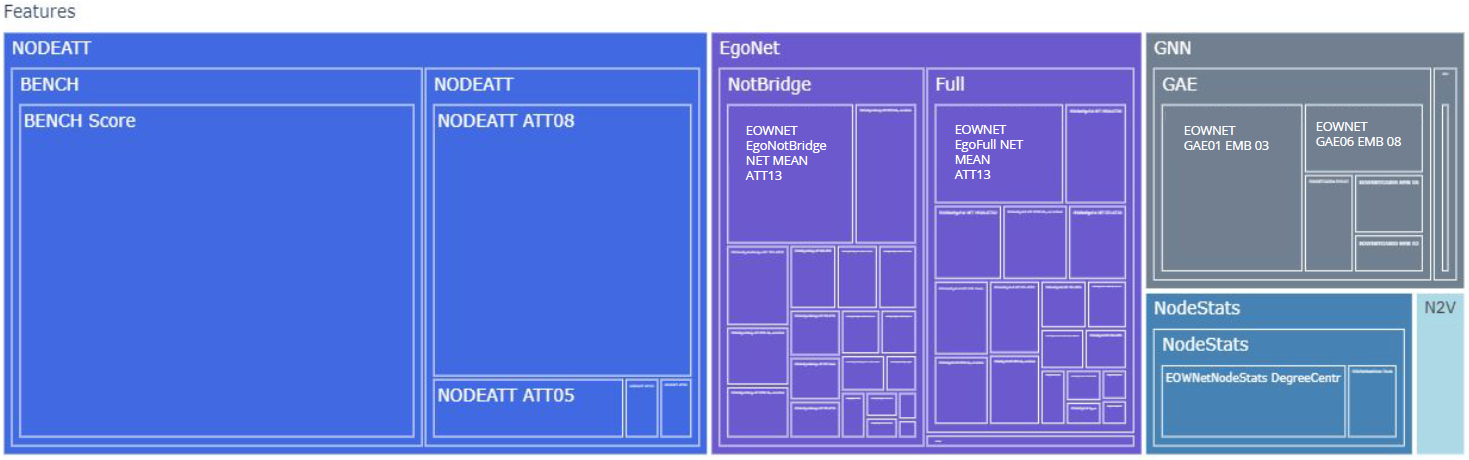}} 
  \end{tabular}
\end{figure}

To quantify the usefulness, impact and importance of the different feature sets on the output model, Figure \ref{fig:PJ_SHAP02} presents a Treemap based on the average of the absolute values of the SHAP values; the complete list of the model features is displayed, and the different colors indicate that they belong to different feature sets. 

In Figure \ref{fig:PJ_SHAP02}(a), it is shown that the feature set $X_{EgoNet}$ (\textbf{D}) contributes, in Application Scoring, 60\% of the model's overall impact. The feature set $X_{GNN,N2V}$ (\textbf{E}) contributes 21\%, of which 19\% correspond to GNN features, while 2\% correspond to Node2Vec. The low importance of Node2Vec features is likely the reason for the limited research on Node2Vec to enhance the prediction of creditworthiness. 

However, in Business Behavioral Scoring, the traditional characteristics now represent 48 \% of the total impact of the model. In contrast, in Business Application Scoring, they represent only 16\% (See Figure \ref{fig:PJ_SHAP02}(b)). Indeed, the BenchScore attribute alone represents 29\% of the total impact. The feature set  $X_{EgoNet}$ (\textbf{D}) represents 30\% of the total impact, the average BenchScore of the ego network being the most relevant attribute.

\subsubsection{Personal Credit Scoring }

In Personal Scoring, the person's characteristics ($X_{Node} + X_{BenchScore}$) produce a more meaningful impact than the business score. The person's attributes represent 37\% and 47\% of the total impact for Application and Behavioral Scoring, respectively. 

\begin{figure}[ht]
  \centering
  \noindent
  \caption{Personal Credit Scoring:  Feature Importance }\label{fig:PN_SHAP01}
  \begin{tabular}{cc}
    \subfigure[Application: Average impact on model output]{\includegraphics[width=0.45\textwidth]{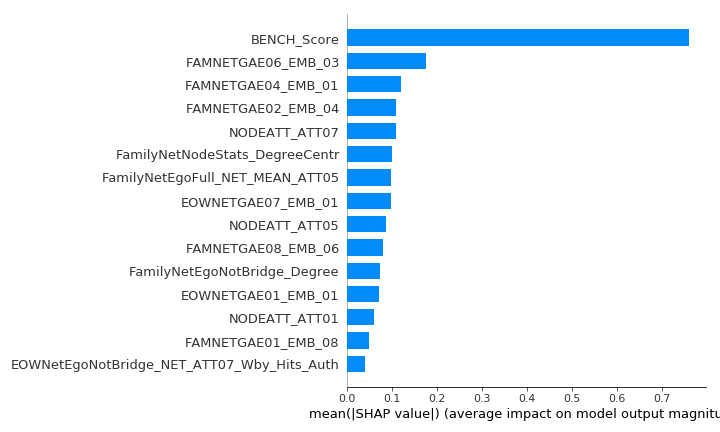}}&
    \subfigure[Application: Impact on model output]{\includegraphics[width=0.45\textwidth]{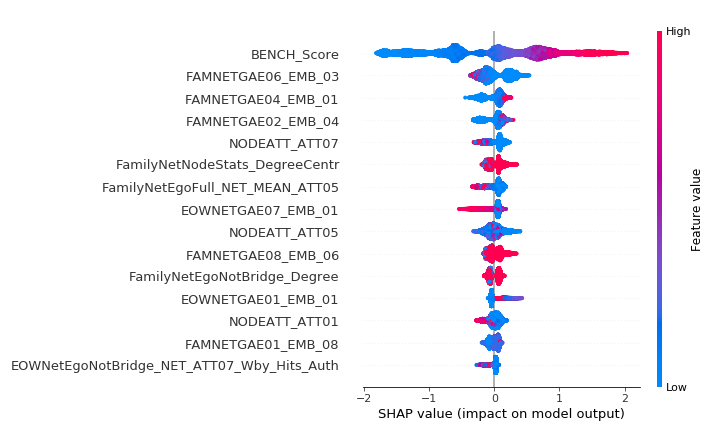}} 
    \\
    \subfigure[Behavioral: Average impact on model output]{\includegraphics[width=0.45\textwidth]{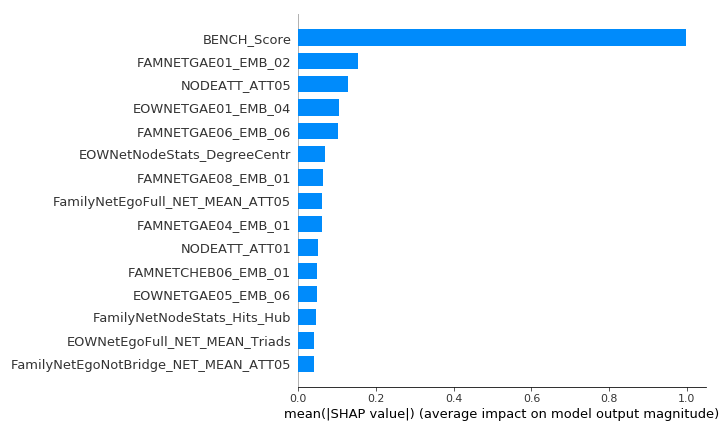}}&
    \subfigure[Behavioral: Impact on model output]{\includegraphics[width=0.45\textwidth]{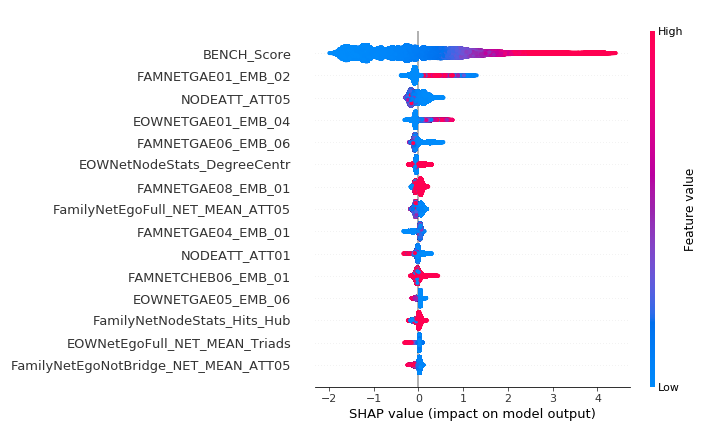}} 
    
  \end{tabular}
\end{figure}

Besides the BenchScore, other relevant features are the amount of  consumer debt (NODEATT\_07) and amount of unused revolving credit, and total debt amount (NODEATT\_01) (see Figures \ref{fig:PN_SHAP01}(a) and  \ref{fig:PN_SHAP01}(c)).

\begin{figure}[ht]
  \centering
  \noindent
  \caption{Personal Credit Scoring: Treemap of Feature Importance, the average impact on model output}\label{fig:PN_SHAP02}
  \begin{tabular}{c}
    \subfigure[Application Scoring]{\includegraphics[width=0.85\textwidth]{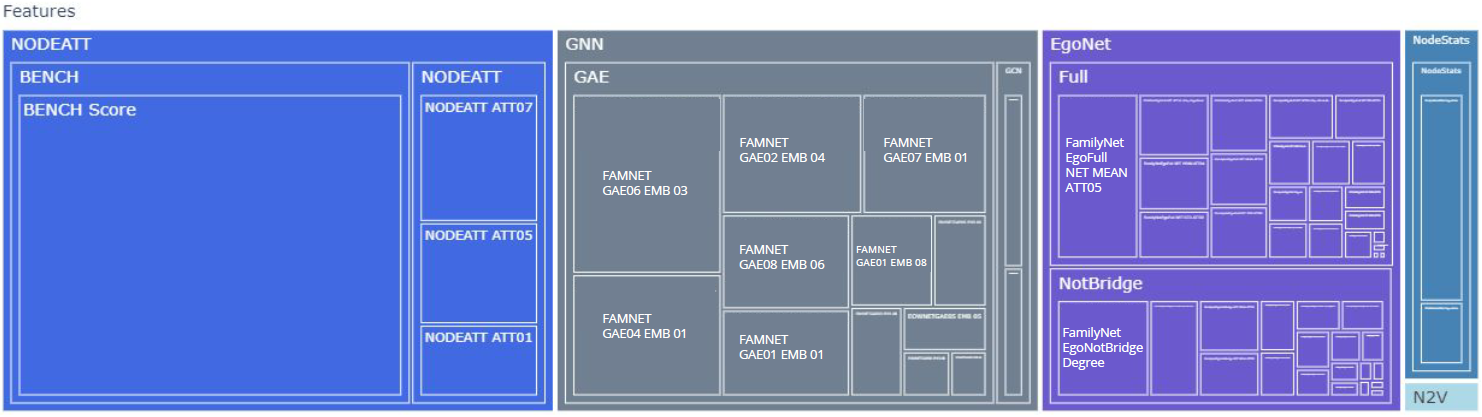}}\\
    \subfigure[Behavioral Scoring]{\includegraphics[width=0.85\textwidth]{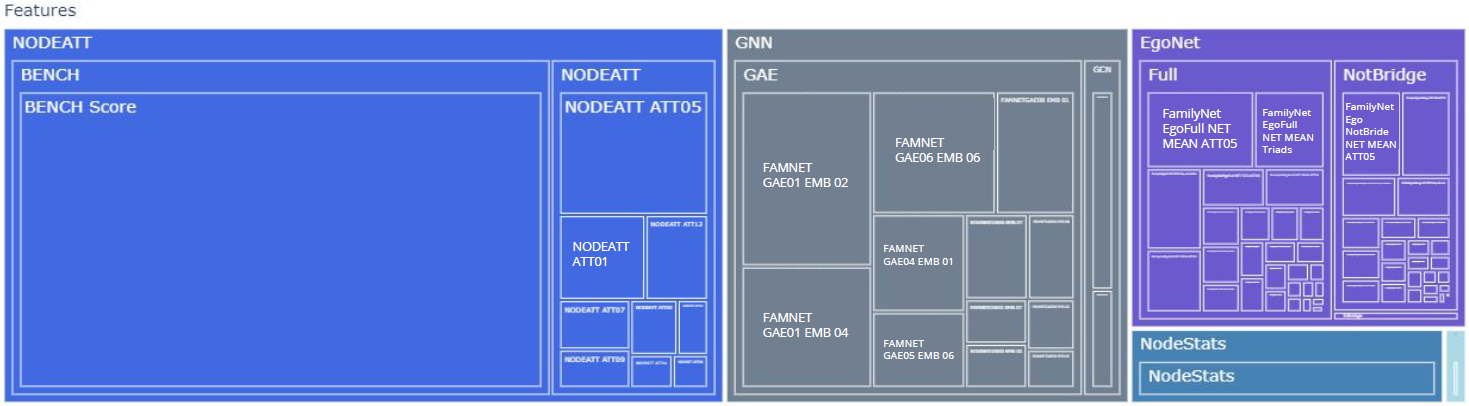}} 
  \end{tabular}
\end{figure}

The combined network features also play an essential role in the final score; the feature sets $X_{EgoNet}$ (\textbf{D})  and  $X_{GNN,N2V}$ (\textbf{E}) represent 25\% and 33.4\%  in Application Scoring (See Figure \ref{fig:PN_SHAP02}(a)), while the impact in Behavioral Scoring (See Figure \ref{fig:PN_SHAP02}(b)) are 18\% and 28.3\% respectively. In both cases, the contribution of Node2Vec features is negligible. The network feature with the highest impact is, as the average neighborhood’s amount, the network's amount of unused revolving credit.

Personal Credit Scoring includes attributes generated with both FamilyNet and EOWNet networks. When analyzing the network-related features, almost all of them, in both scenarios, are FamilyNet features. These results show us both the suitability of the network used to characterize borrowers and the importance of the type of relationship used to build the network. In this study, family ties are the most appropriate to characterize borrowers as regards the problem of individual credit scoring.

\section{Conclusions}\label{sec:Conclusions}
This study presents an information processing methodology that allows us to assess the additional value of social-interaction data to approach the credit scoring problem for thin-file clients. This framework is applied in four scenarios arising from the consideration of  all combinations of Application and Behavioral Scoring of individual and business lending. Additionally, this methodology allows the evaluation of different GRL approaches to feature extraction from social networks: hand-crafted feature engineering, Node2Vec, and Graph Neural Networks.
The results show an improvement in creditworthiness assessment performance when different GRL approaches are combined. Specifically, two of the three GRL methods significantly enhance credit scoring models, namely the Hand-crafted Feature Engineering and Graph Neural Networks, which have the greatest impact  when used together. We believe this to be very relevant to the community because, until now, these two methods have been used independently. On the other hand, we have found that the contribution of Node2Vec is negligible. This result seems to justify the limited research conducted on Node2Vec as a feature-engineering method for credit scoring.

As a baseline, we use a credit scoring model developed by a financial institution. This model, the BenchScore, already outperforms the credit bureau model they obtain from the credit bureau offices, and our methodology allows us to obtain better results in each of the four scenarios.

The highest value of the proposed approach is found in Unbanked Application Scoring. The unbanked applicants, individuals, and companies, lack behavioral information, which, as it turns out, is one of the best predictors of creditworthiness. Our approach overcomes the lack of behavioral information and delivers a proper credit risk assessment using graph data. In this way, applicants have greater access to the financial system. In the case of the Behavior Scoring models, our methodology also improves performance. In both cases, the maximum improvement in predictive performance is achieved when these GRL methods are used together.

Explanatory measures, such as SHAP values, allow us to understand each attribute's contribution. If the impact on the output model is measured in this way, the baseline model (BenchScore), although it continues to be an essential attribute, has a diminished effect because it is in the presence of other good predictors. This feature importance analysis allows us to understand that we cannot solely examine a company's characteristics to evaluate the company, especially if it is unbanked. We also have to understand that they are part of an ecosystem in which the owners, suppliers, clients, and related companies are essential. The business ecosystem information allows us to improve the creditworthiness assessment performance. A similar situation occurs in Personal Credit Scoring, although with less intensity. The network data allows us to address the scarcity of information and achieve a better credit risk assessment.

Our research shows that there is still room for improvement in incorporating network information into the credit scoring problem. This methodology goes in the right direction, improving the performance of creditworthiness assessment, and it has great value for unbanked and under-banked people and even in the management of  portfolio's credit risk.

\section*{Acknowledgments}

This work would not have been accomplished without the financial support of CONICYT-PFCHA / DOCTORADO BECAS CHILE / 2019-21190345. The second author acknowledges the support of the Natural Sciences and Engineering Research Council of Canada (NSERC) [Discovery Grant RGPIN-2020-07114]. This research was undertaken, in part, thanks to funding from the Canada Research Chairs program.
The last author thanks the partial support of  FEDER funds through the MINECO project TIN2017-85827-P and the European Union’s Horizon 2020 research and innovation program under the Marie Sklodowska-Curie grant agreement No 777720.


\bibliographystyle{apacite}

\newpage
\appendix

\setcounter{table}{0}
\renewcommand{\thetable}{A\arabic{table}}

\section{Feature Description}

\begin{table}[ht]
\footnotesize
\centering
\caption{Taxonomy of features used in the experiments  and nomenclature that we used for the management of the data in the experiments.}
\label{table:FeatureNomenclature}
\resizebox{0.75\textwidth}{!}{%
\begin{tabular}{|m{2.5cm}|m{3cm}|cm{6cm}|}
\hline
{\parbox{2.5cm}{\centering Feature Set}}&{\parbox{3cm}{\centering Nomenclature}}& {\parbox{2.5cm}{\centering Prefix/suffix}}& {\parbox{6cm}{\centering Description}} \\ \hline

\multirow{5}{*}{\parbox{2.5cm}{\centering \textbf{Node Features} $(X_{Node})$}} & {\parbox{3cm}{\centering Identifier}} & $NODEATT$& Feature subset identifier  \\ \cline{2-4}
 &  \multirow{4}{*}{\parbox{3cm}{\centering Borrower feature identifier}} &$ATT01, \cdots , ATT04$ &  \parbox{6cm}{The debt situation characterized by the delinquency level} \\ 
 &&$ATT05, \cdots , ATT08$& \parbox{6cm}{ The debt type: revolving, consumer, commercial, or mortgage }\\ 
 &&$ATT09, \cdots , ATT13$& \parbox{6cm}{ Other aspects of the customer's debt, payments in arrears, and the time in the financial system }\\ 
 &&$Bench\_Score$& The benchmark score \\ \hline

\multirow{9}{*}{\parbox{2.5cm}{\centering \textbf{Node Statistics} $(X_{NodeStats})$}} & {\parbox{3cm}{\centering Identifier}} & NodeStats &Feature subset identifier   \\ \cline{2-4}
  &  \multirow{6}{*}{\parbox{3cm}{\centering Statistic identifier}} &$DegreeCentr$ & Degree centrality \\ 
  &&$Triads$ &   Number of triads \\ 
  &&$PageRank$ & PageRank Algorithm \\ 
  &&$ArtPoint$ & Articulation point\\ 
  &&$Hits\_Auth$ & Hits algorithm Authority score\\ 
 &&$Hits\_Hub$ & Hits algorithm Hub score\\ \cline{2-4}
  &  \multirow{2}{*}{\parbox{3cm}{\centering Network identifier}} & EOWNET & EOWNet Network  \\ 
  && FamilyNet &   Family Network \\ \hline

\multirow{8}{*}{\parbox{2.5cm}{\centering \textbf{EgoNetwork Agreggation Features} $(X_{EgoNet})$}} & {\parbox{3cm}{\centering Borrower feature identifier}} & $ATT01, \cdots , ATT13$ & Borrower Feature  \\ \cline{2-4}

  &  \multirow{2}{*}{\parbox{3cm}{\centering Network identifier}} & EOWNET & EOWNet Network  \\
  && FamilyNet &   Family Network \\ \cline{2-4}
  
  &  \multirow{2}{*}{\parbox{3cm}{\centering Aggregation Function}} & MEAN & Mean  \\
  && STD &   Standard Deviation \\ \cline{2-4}
  
   &  \multirow{3}{*}{\parbox{3cm}{\centering Edges}} & Full & All edges  \\
  && NotBridge &  Edges that are not bridges \\
  && IsBridge  &   Edges that are bridges \\ \cline{2-4} 
   & {\parbox{3cm}{\centering Weighted Aggregations}} & $Wby+Feature$ &  Suffix for weighted aggregations \\ \hline

 \multirow{4}{*}{\parbox{2.5cm}{\centering \textbf{Node2Vec Features} $(X_{EgoNet})$}} & {\parbox{3cm}{\centering Identifier}} & N2V &Feature subset identifier   \\ \cline{2-4} 
   & {\parbox{3cm}{\centering  Embedding Identifier}}& $EMB\_01, \cdots , EMB\_08$ &  The embedding number \\ \cline{2-4} 
     &  \multirow{2}{*}{\parbox{3cm}{\centering Network identifier}} & EOWNET & EOWNet Network  \\
  && FamilyNet &   Family Network \\ \hline

 \multirow{6}{*}{\parbox{2.5cm}{\centering \textbf{Graph Neural Network Features} $(X_{GNN})$}} & \multirow{2}{*}{\parbox{3cm}{\centering GNN Identifier}} & CHEB & Graph Convolutional Network (GCN)  \\ 
  & & GAE &  Graph Autoencoder \\ \cline{2-4} 
  & \parbox{3cm}{\centering Borrower feature identifier} & $01, \cdots , 13$ & Borrower Feature   \\ \cline{2-4}
   &{\parbox{3cm}{\centering  Embedding Identifier}}& $EMB\_01, \cdots , EMB\_n$ &  The embedding number, where $n=3$ for CHEB and $n=8$ for GAE \\ \cline{2-4} 
     &  \multirow{2}{*}{\parbox{3cm}{\centering Network identifier}} & EOWNET & EOWNet Network  \\
  && FAMNET &   Family Network \\ \hline
\end{tabular}}
\end{table}

\end{document}